\documentclass[twocolumn]{aastex62}
\usepackage[percent]{overpic}
\usepackage{amsmath}
\usepackage{hyperref}

\newcommand{\um}{$\mu$m}
\usepackage{float}

\shorttitle{X-ray Emission of IMPS}
\shortauthors{Nu\~{n}ez et al.}

\begin{document}
\turnoffedit1

\title{Characterizing the X-ray Emission of Intermediate-Mass Pre-Main-Sequence Stars}

\correspondingauthor{Evan Haze Nu\~{n}ez}
\email{enunez@astro.caltech.edu}

\author[0000-0001-5595-757X]{Evan H. Nu\~{n}ez}
\affil{Department of Physics \& Astronomy, California State Polytechnic University, 3801 W. Temple Ave, Pomona, CA 91671, USA}
\affil{California Institute of Technology, 1200 E. California Blvd., MC 249-17, Pasadena, CA 91125, USA}

\author{Matthew S. Povich}
\affil{Department of Physics \& Astronomy, California State Polytechnic University, 3801 W. Temple Ave, Pomona, CA 91671, USA}
\affil{California Institute of Technology, 1200 E. California Blvd., MC 249-17, Pasadena, CA 91125, USA}

\author{Breanna A. Binder}
\affil{Department of Physics \& Astronomy, California State Polytechnic University, 3801 W. Temple Ave, Pomona, CA 91671, USA}

\author{Leisa K. Townsley}
\affil{Department of Astronomy \& Astrophysics, Pennsylvania State University, 525 Davey Laboratory, University Park, PA 16802, USA}

\author{Patrick S. Broos}
\affil{Department of Astronomy \& Astrophysics, Pennsylvania State University, 525 Davey Laboratory, University Park, PA 16802, USA}


\begin{abstract}
We use X-ray and infrared observations to study the properties of three classes of young stars in the Carina Nebula:  intermediate-mass \edit1{(2--5~M$_\odot$)} pre-main sequence stars (IMPS\edit1{; i.e. intermediate-mass T Tauri stars}), late-B and A stars on the zero-age main sequence (AB), and lower-mass T Tauri stars (TTS). We divide our sources among these three sub-classifications and further identify disk-bearing young stellar objects versus diskless sources with no detectable infrared (IR) excess emission using IR (1--8 $\mu$m) spectral energy distribution modeling. We then perform X-ray spectral fitting to determine the hydrogen absorbing column density ($N_{\rm H}$), absorption-corrected X-ray luminosity ($L_{\rm X}$), and coronal plasma temperature ($kT$) for each source. We find that the X-ray spectra of both IMPS and TTS are characterized by similar $kT$ and $N_{\rm H}$, and on average $L_{\rm X}$/$L_{\rm bol} \sim4\times10^{-4}$. IMPS are systematically more luminous in X-rays (by $\sim$0.3 dex) than all other sub-classifications, with median $L_{\rm X} = 2.5\times10^{31}$ erg s$^{-1}$, while AB stars of similar masses have X-ray emission consistent with TTS companions. These lines of evidence converge on a magneto-coronal flaring source for IMPS X-ray emission, a scaled-up version of the TTS emission mechanism. IMPS therefore provide powerful probes of isochronal ages for the first $\sim$10 Myr in the evolution of a massive stellar population, because their intrinsic, coronal X-ray emission decays rapidly after they commence evolving along radiative tracks. We suggest that the most luminous (in both X-rays and IR) IMPS could be used to place empirical constraints on the location of the intermediate-mass stellar birth line.
\end{abstract}


\section{Introduction} \label{sec_intro}

The observed X-ray emission of low-mass, T Tauri stars (TTS) is primarily coronal in origin \citep[e.g.,][]{preibisch_2005,stassun_2006,stassun_2007,Telleschi07}. TTS coronae are roughly an order of magnitude hotter than the solar corona, and more X-ray luminous than the Sun by up to a factor of $\sim10^5$ \citep[][herafter P05]{preibisch_2005}. 
 P05 derived $L_{\rm X}$--$M_\star$ and $L_{\rm X}$--$L_{\rm bol}$ relationships for low-mass pre-main-sequence (PMS) stars (0.1--2~$M_{\odot}$) observed as part of the {\em Chandra} Orion Ultradeep Project \citep[COUP;][]{COUP}. Among the 870 X-ray detected stars in the COUP optical sample (a subset of stellar members from \citealp{hillenbrand97}), only 20 (${\sim}2\%$) low-mass stars ($M<2~M_{\odot}$) produced total-band (0.5--8~keV) $\log{L_X} >31$~erg s$^{-1}$, but the majority of intermediate-mass stars (2--3~$M_\odot$) exceeded this luminosity threshold (see Fig.~3 of P05).

High-quality X-ray sources in the COUP sample were generally well-fit by two-temperature thermal plasma models, with a characteristic softer component at $T_1\approx 10$~MK and a harder component ranging over 20~MK $\la T_2 \la 60$~MK, with $T_2$ a steeply increasing function of observed X-ray flux (P05). While the actual distribution of plasma temperatures within coronally-active stars is certainly more complicated and varied \citep{gudel+07}, a simple 2-component thermal plasma model is generally adequate to fit the observed spectra of faint X-ray sources \citep{getman+10}. \edit1{In stars with active mass accretion from disks, accretion shocks may contribute to the soft X-ray spectrum at $kT < 1$~keV \citep{gunther+2007}.} 

Coronal TTS X-ray emission indicates the presence of a strong surface magnetic field \citep{stassun_2006,stassun_2007}, with typical mean field strengths 1-5 kG \citep{alecian_2019,Sokal20}. The dynamos producing these magnetic fields are driven by convective interior structure, typical of low-mass TTS evolving along Hayashi tracks. In the case of intermediate-mass, PMS stars (IMPS; 2--4~$M_\sun$)
%
the development of a radiative interior marks the beginning of a rapid decline in observed X-ray emission \citep{mayne+07,mayne_10,gregory_2016,getman+2021}. 
%
IMPS with spectral types G through early K are the progenitors of zero-age main sequence (ZAMS) late-B through A-type stars. Fully-radiative AB stars possess neither magnetic dynamos nor sufficiently strong winds to power the shock-driven X-rays 
\edit1{commonly observed from massive, O- and early B-type stars \citep{gagne_2011,naze+11,preibisch+2021}.} 
This implies that any intrinsic X-ray emission from convective IMPS should disappear over timescales shorter than the ${<}10$~Myr ZAMS arrival times of ${>2}~M_{\sun}$ stars \citep{siess_2000,haemmerle_2019}.

The stellar initial mass function (IMF), coupled with shorter PMS evolutionary timescales, makes IMPS rare compared to lower-mass TTS, hence statistically robust samples of X-ray emitting IMPS can only be studied in very young, very massive star-forming regions. 
The \textit{Chandra} Carina Complex Project \citep[CCCP;][]{townsley_2011} covered 1.42 deg$^2$ of the Great Nebula in Carina with the \textit{Chandra X-ray Observatory}, revealing ${>}10,000$ young stellar members as X-ray point sources \citep{broos_2011,CCCP_Class}.
As part of CCCP, \citet{povich_2011} analyzed the {\em Spitzer Space Telescope} point-source population to produce a Pan-Carina YSO catalog of 1432 young stellar objects with mid-infrared excess emission from circumstellar disks and infalling envelopes, 410 of which were also detected in X-rays. Based on the spectral energy distribution (SED) modeling of these predominantly intermediate-mass YSOs, \citet{povich_2011} found that X-ray detection correlated with cooler stellar photospheres and higher disk masses, both indicators of earlier evolutionary ages. These results provided the first indication that convection-driven, magneto-coronal X-ray emission was present in the Carina IMPS population. 
\citet[][hereafter P19]{povich_2019} presented a detailed study of 2269 CCCP X-ray sources whose {\em Spitzer} counterparts lacked detectable 4.5~\um\ excess emission to constrain the duration of star formation in various sub-regions within the complex.  The fraction of X-ray detected, intermediate-mass stars is lower in more evolved populations (the Tr 15 cluster and distributed populations, 6--10~Myr) compared to the younger clusters (Tr 14 and 16, ${<}3$~Myr). The highly time-dependent nature of IMPS X-ray emission can provide new insights into both stellar properties and evolutionary timescales within young, massive star-forming regions.

In this paper we fit thermal plasma models to 370 of the brightest CCCP X-ray sources, excluding those associated with known or candidate OB stars \citep{gagne_2011,povich_2011,alexander_2016}. 
Using the novel IR SED modeling technique developed by P19, we classify sources as IMPS (including a special sub-category of partially-radiative, R-IMPS), TTS, late-B and A-type (AB) stars on the ZAMS, or unclassified sources that could not be confidently placed into any one of the former categories.
We investigate trends in X-ray emission properties, including thermal plasma temperature, hydrogen absorbing column density, and X-ray luminosity, both within and across these classes. We also analyze both the X-ray and IR properties to investigate the $L_X$--$L_{\rm bol}$ and  $L_X$--$M_{\star}$ relationships across the often-neglected 2--4~$M_{\sun}$ mass range. \deleted{ stellar mass and X-ray luminosity relation. With these approaches we are able to test all of the hypothesis mentioned earlier.}

This paper is structured as follows. In Section \ref{sec_sample} we describe our data, source selection criteria, and the IR SED and X-ray spectral model fitting. 
In Section \ref{sec_analysis} we analyze physical properties from the X-ray spectral analysis results. In Section \ref{sec_discussion} we discuss trends across classifications, $L_X$--$L_{\rm bol}$ and  $L_X$--$M_{\star}$ relations, and several interesting individual IMPS. In Section \ref{sec_conclusions} we summarize our conclusions.

\section{Sample Selection} \label{sec_sample}
\subsection{IR and X-ray Observations} \label{sec_observations}

Our study sample is drawn from 410 YSOs and 3657 ``diskless" (no excess emission above a stellar photosphere at 4.5~\um) stars positionally matched to X-ray point-sources in the CCCP catalog \citep{broos_2011,povich_2011,povich_2019}.
The \textit{Spitzer Space Telescope} Vela-Carina Survey Point Source Archive provided mid-infrared (MIR) photometry from IRAC \citep[at 3.6, 4.8, 5.6, and 8.0 $\mu$m;][]{IRAC}, as well as near-infrared (NIR)  \textit{JHK$_S$} photometry from the Two-Micron All Sky Survey (2MASS) Point Source Catalog \citep{skrutskie_2006}. The Vela-Carina data have $2\arcsec$ spatial resolution and typical sensitivity of $[4.5]\la 15.5$~mag (less sensitive in regions of bright MIR nebulosity). For YSOs, \citet{povich_2011} additionally provided \textit{Spitzer}/MIPS 24~\um\ photometry detections or upper limits.
CCCP X-ray point sources were observed with the \textit{Chandra}/ACIS-I detector, \citep[][]{garmire_2003}, with positional, photometric, temporal, and spectral information obtained using \textit{ACIS Extract} \citep{broos_2010}. We restrict our sample to the brightest CCCP X-ray sources, those with ${\ge}50$ net counts (in the full ACIS 0.5--8 keV energy band) and ${\ge}5\sigma$ significance to ensure reliable spectral analysis. There were 1,049 sources from the CCCP sample that fit our X-ray selection criteria.
\subsection{Cleaning the Initial Sample}\label{sec_decontaminant}

Not all of the sources in our initial sample were genuine low- or intermediate-mass young stars in the Carina Nebula.
Our selection criteria may include massive, OB members of the Carina Nebula and contaminating foreground stars.

To identify residual foreground stars that were not previously flagged as contaminants by \citet{CCCP_Class}, P19 used parallax information from Gaia DR2 \citep{gaia_collab_dr2} for all available sources to distinguish between members and non-members of Carina. There were 19 sources in our initial X-ray bright sample with parallaxes that were consistent with being foreground sources; we excluded these.

We also cross-referenced catalogs of spectroscopically confirmed OB stars \citep{gagne_2011,alexander_2016,damiani_2017} and removed 78 known massive stars from our sample.  

\subsection{IR Source Classification from SED Modeling} \label{sec_src_class}

Following \citet{povich_2011,povich_2019}, we divide the MIR counterparts of X-ray sources into two broad categories: YSOs with excess 4.5 $\mu$m emission consistent with circumstellar disks (usually coupled with excess 5.8 $\mu$m and 8.0 $\mu$m emission), and stars with no excess emission at 4.5 $\mu$m and hence no warm dust disks. These ``diskless" stars may have (1) 5.8 and/or 8.0~\um\ detections consistent with a Rayleigh-Jeans spectrum, (2) no photometric detection at wavelengths longer than 4.5~\um, or (3) marginal IR excess emission at 5.8 $\mu$m or 8.0 $\mu$m. In cases (2) or (3) the stars could still possess disks with large inner holes, and in all three cases the stars could still possess debris disks that would produce excess emission at ${>}10$~\um\ wavelengths. These classifications were determined by fitting model spectral energy distributions (SEDs) to the 1--8~\um\ broadband photometry of all sources.

 Diskless SEDs were fit with a set of 100,000 ``naked'' stellar photosphere PMS models (P19) using the \citet{robitaille_2007} SED fitting tool. These models sample a range in stellar mass and age ($M_\star$ and $t_\star$), then convert these properties to the corresponding photospheric temperature and radius ($T_{\rm eff}$ and $R_{\rm eff}$) using PMS evolutionary tracks \citep{siess_2000,bernasconi_1996}.
For SEDs showing IR excess (the minority of our sample), we use the \citet{robitaille_2006} YSO model fit parameters published by \citet{povich_2011}.
For both YSOs and diskless SED models, we  assumed the \citet{indebetouw_2005} extinction curve, which introduced a new free parameter, $A_{\rm V}$, allowed to vary from 0 to 15~mag (P19). Sources for which no SED models provided satisfactory fits according to the $\chi^2$ criteria of \citet{povich_2011} or P19 were excluded from our sample. 


Using the methodology developed by P19, the likelihood of each model fit to a given SED is weighted individually using two distinct, mass- and age-dependent weighting functions, one based on the disk destruction timescale \citep[see][and references therein]{povich_2016} and the other on the X-ray emission decay timescale for PMS stars from \citet[][hereafter G16]{gregory_2016}. The result is a set of probability distributions for the model stellar parameters $M_\star$, $t_\star$, $T_{\rm eff}$, and $L_{\rm bol}$. 

We classify each source into one of four temperature--mass ($TM$) classifications: IMPS, TTS, AB stars (A and late B-type stars on or near the ZAMS in the same mass range as IMPS, including Herbig Ae/Be stars), or massive OB stars. 
We combine the weighted $T_{\rm eff}$ and $M_\star$ probability distributions for each source into a two dimensional weighted $TM$ probability distribution ($P_i$) to determine the relative probability of each source being any particular $TM$ class. 
Preliminary $TM$ Class Distributions are summed to a single value that we define as Preliminary $TM$ Class probabilities ($P_{TM}$) to compare the relative likelihood of all Preliminary $TM$ Classes to one another. We define the $P_{TM}$ pairings using 
\begin{equation} \label{eq_TM_dist}
\begin{split}
    P_{TM} = \\
    \begin{cases}
    P_{IMPS};   &   \sum P_i(T_{\rm eff} \leq 7300~{\rm K}, \; 2~M_{\odot}\le M_{\star} < 8~M_{\odot}) \\
    P_{TTS};    &   \sum P_i(T_{\rm eff} \leq 7300~{\rm K}, \; M_{\star} < 2~M_{\odot})  \\
    P_{AB};     &   \sum P_i(T_{\rm eff} > 7300~{\rm K}, \;  2~M_{\odot}\le M_{\star} < 8~M_{\odot})   \\
    P_{OB};     &   \sum P_i (M_{\star} \ge 8~M_{\odot}).
    \end{cases}
\end{split} 
\end{equation}

To make the final classifications, we compare the two most probable preliminary $TM$ class probabilities to one another, according to
\begin{equation} \label{eq_TM_class}
    {\rm TM~Class} = 
    \begin{cases}
    TM1; &  P_{TM1} > 2P_{TM2} \\
    {\rm U}; & \text{otherwise}, \\
    \end{cases}
\end{equation}
where $P_{TM1}$ (and $P_{TM2}$) are the preliminary $TM$ class probabilities defined in Equation \ref{eq_TM_dist} for the first- (TM1) and second-most (TM2) probable $TM$-Classes for an individual SED.
``U'' designations are given to sources that remain unclassified based on this criterion. Sources that were classified as OB stars, 30 in total, were removed from our sample. 



\begin{figure*}[thp]
    \centering
        \begin{overpic}[scale=0.4]{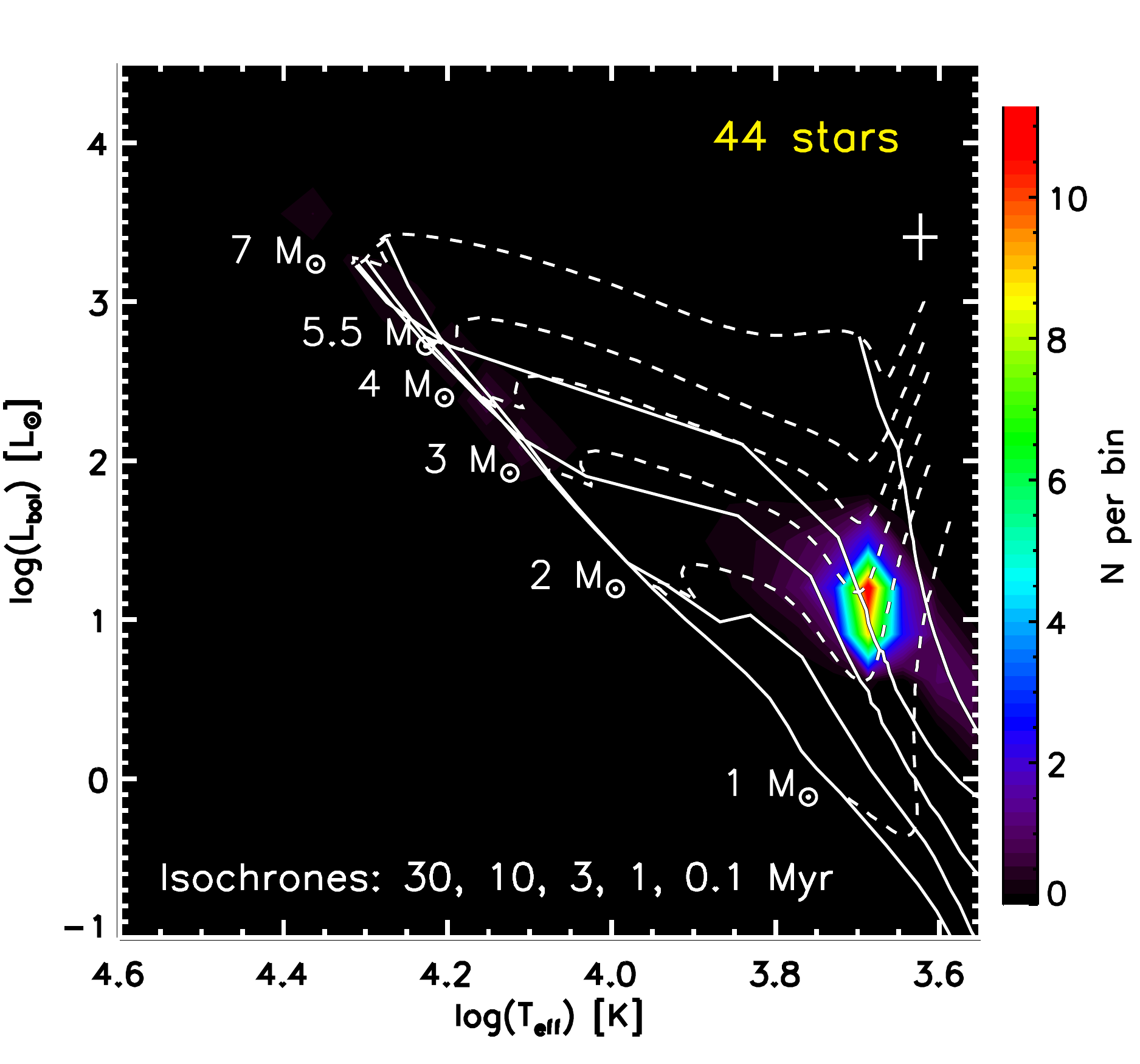} \put(15,80){\color{yellow}\large IMPS} 
        \end{overpic}
        \begin{overpic}[scale=0.4]{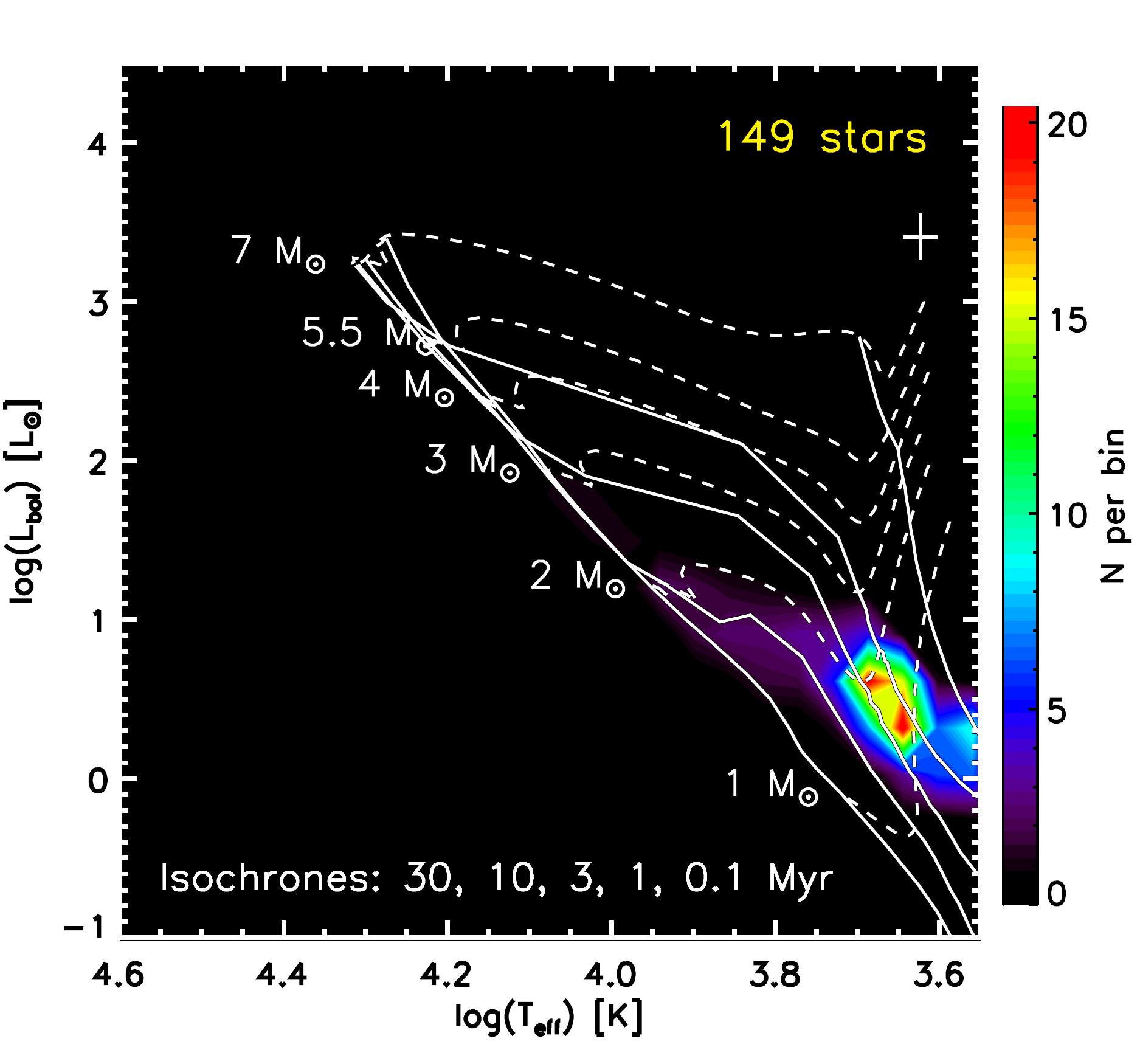} \put(15,80){\color{yellow}\large TTS} 
        \end{overpic}
        \begin{overpic}[scale=0.4]{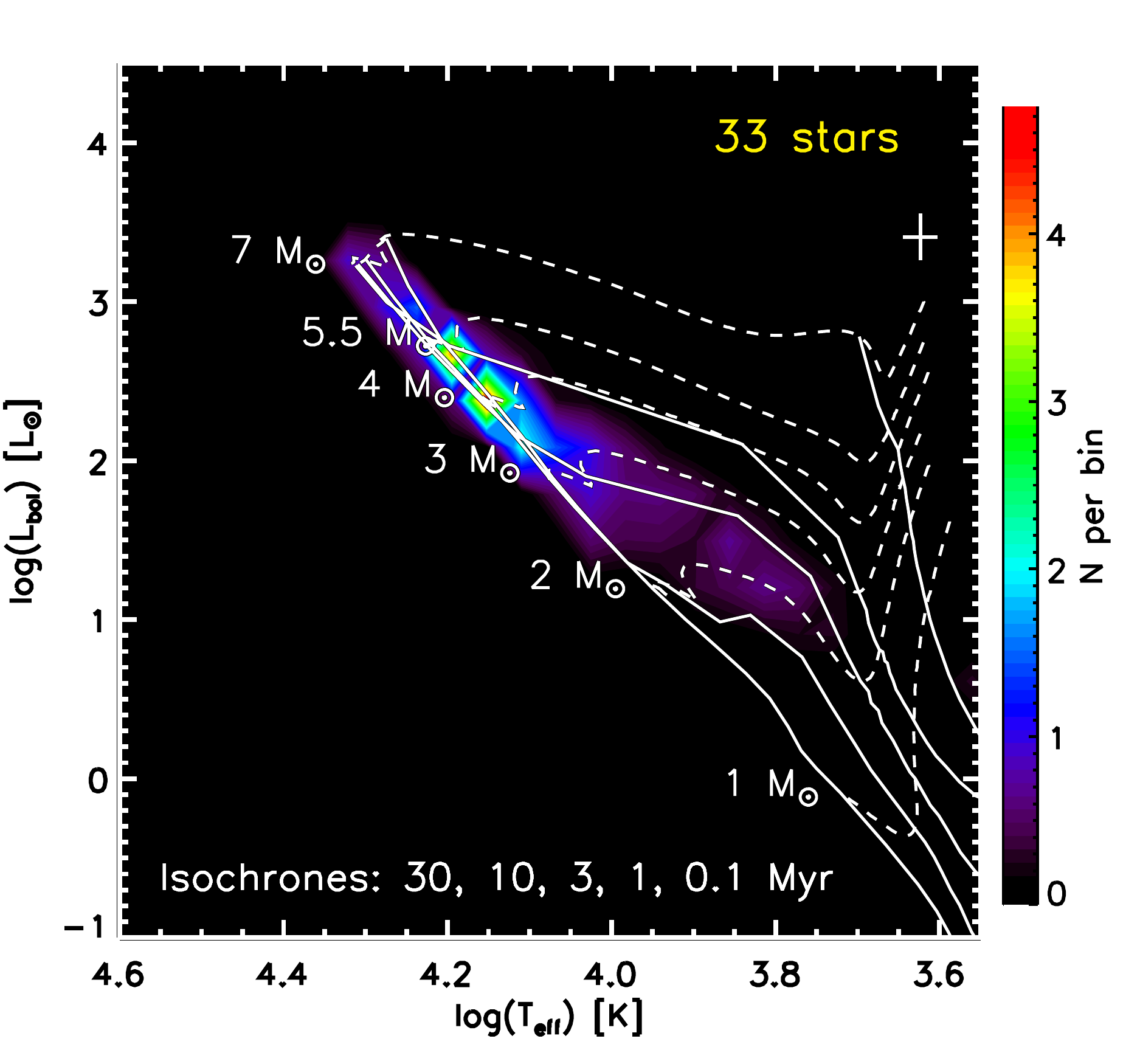} \put(15,80){\color{yellow}\large AB} 
        \end{overpic}
        \begin{overpic}[scale=0.4]{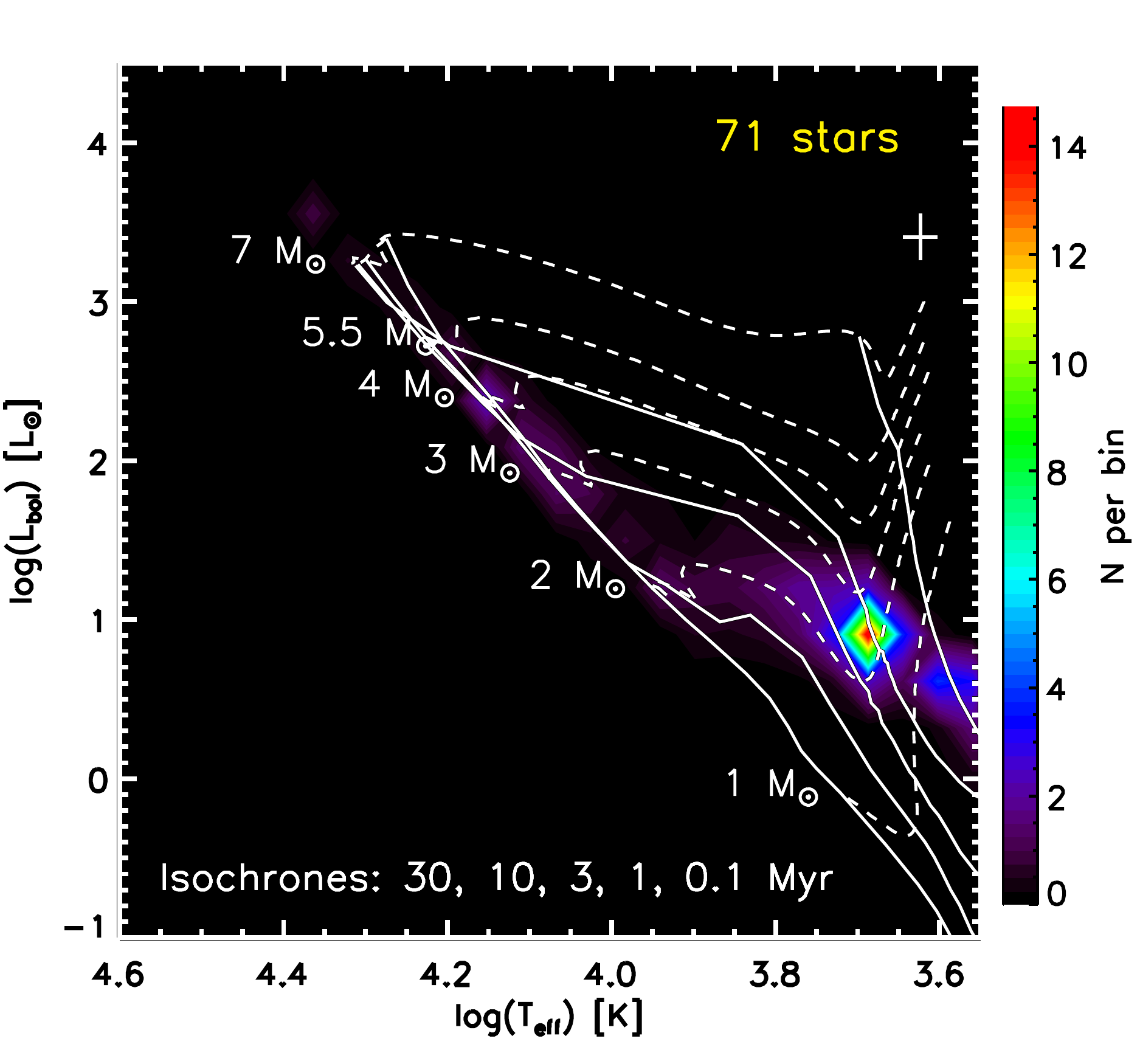} \put(15,80){\color{yellow}\large U} 
        \end{overpic}
    \caption{Composite pHRDs for diskless, X-ray bright sources (Section~\ref{sec_src_class}), divided by $TM$ class. IMPS fall within the 2--4 $M_{\odot}$ mass range and have ages ${\sim}1$~Myr; diskless TTS fall in the  1--2~$M_{\odot}$ range with ages between 1--3~Myr; AB populate the 3--7 $M_{\odot}$ ZAMS; and U sources are primarily borderline cases between IMPS and TTS with masses ${\sim}2~M_{\odot}$.
    \citet{siess_2000} isochrones and evolutionary tracks are shown as solid and dashed lines, respectively. The cross at the top right of each panel shows the bin size for each axis.
    \label{fig_phrd}
    }
\end{figure*}
To visualize these $TM$ Classes we constructed Probabilistic H-R Diagrams (pHRDs, see P19) by summing the probability distributions for all sources in a given $TM$ Class. The pHRDs for three $TM$ Classes plus unclassified are plotted as two-dimensional histograms in the four panels of Figure \ref{fig_phrd} (the OB class is excluded and would extend above the maximum $L_{\rm bol}$ plotted here). These pHRDs demonstrate that even the TTS sources in our sample are typically more massive than the Sun, which reflects both the relatively shallow IR photometry and our selection of the brightest CCCP X-ray sources. U sources consist primarily of stars whose SED model fits straddled the ${\sim}2~M_{\odot}$ mass cutoff separating IMPS from TTS; based on the IMF the majority of U sources are likely T Tauri stars. 

\edit1{The numerical breakdown of sources assigned to each $TM$ class in the cleaned initial sample are given in Table \ref{tab_src_cnt}, which also summarizes our source counts in the two subsequent sample refinement steps.}

\begin{table}[htb]
\begin{center}
\caption{Sample Refinement} \label{tab_src_cnt}
    \begin{tabular}{cccccc}
    \hline
    \hline
                &$T_{\rm eff}$\tablenotemark{a} &$M_\star$  &Initial   &Reclass &Final   \\
    $TM$ Class  &(K)   &($M_{\odot}$)  &\S\ref{sec_observations}--\ref{sec_src_class}  &\S\ref{sec_ir_damiani} & \S\ref{sec_xspec}     \\
    \hline
    IMPS        &${\leq}7300$                   &2--8       &57         &59     &54     \\
    R-IMPS      &${\geq}5600$\tablenotemark{b}  &2--8       &\nodata    &24     &23     \\
    TTS         &${\leq}7300$                   &${\leq}2$  &194        &196    &176    \\
    AB          &${>}7300$                      &2--8       &52         &39     &35     \\
    U           &\nodata                        &\nodata    &102        &84     &82     \\
    \hline
    All         &\nodata                        &\nodata    &405        &402    &370   \\
    \hline
    \end{tabular}
\end{center}
    \tablenotemark{a}{$T_{\rm eff}=7300$~K is the canonical value for an F0 V star.}\\ 
    \tablenotemark{b}{R-IMPS classifications require spectroscopic $T_{\rm eff,S}$.}
\end{table}

\subsection{Stars with Spectroscopically-Measured Effective Temperatures} \label{sec_ir_damiani}
\citet[][hereafter D17]{damiani_2017} analyzed optical spectra of sources in the Trumpler (Tr)~14 and 16 clusters in Carina obtained as part of the Gaia-ESO survey with the FLAMES/Giraffe multi-fiber spectrometer at the ESO VLT/UT2 telescope. They broadly separated their sources into early-type and late-type stars, and for the latter they reported a spectroscopically-measured effective temperature $T_{\rm eff,S}$ for each star. We compared our samples and found no matches to their early-type stars but found 64 matches to their late-type stars (13 AB, 12 IMPS, 25 TTS, and 14 unclassified). Three matched sources that we had classified as AB had $T_{\rm eff,S}$ $<$ 4800~K; these were removed from our sample as suspected foreground stars. 

We refit the SEDs for the remaining 61 sources, this time constraining the model $T_{\rm eff}$ parameter using the $T_{\rm eff,S}$ and (assumed Gaussian) uncertainty values reported by D17 (see P19 for details). We were thus able to classify 8 previously-unclassified sources as TTS, 2 as IMPS, and define a new $TM$ Class of radiative-IMPS (R-IMPS). R-IMPS are a transitional stage between fully-convective IMPS and fully-radiative AB stars in the 2--4~$M_{\odot}$ range. They have begun to traverse the Henyey tracks and have $7300~{\rm K}>T_{\rm eff,S}> 5300$~K, corresponding to spectral types F and early-G. We hence reclassified 24 sources as R-IMPS, including 10 initially classified as AB sources, 6 as IMPS, and 8 previously unclassified.  \edit1{The numerical breakdown of sources after this step of sample refinement is shown in the ``Reclass'' column of Table \ref{tab_src_cnt}.} 
The large majority of R-IMPS (21 out of 24) exhibited no MIR excess and were fit with diskless PMS models. 

\begin{figure*}[ht]
    \centering
    \includegraphics[scale=0.4]{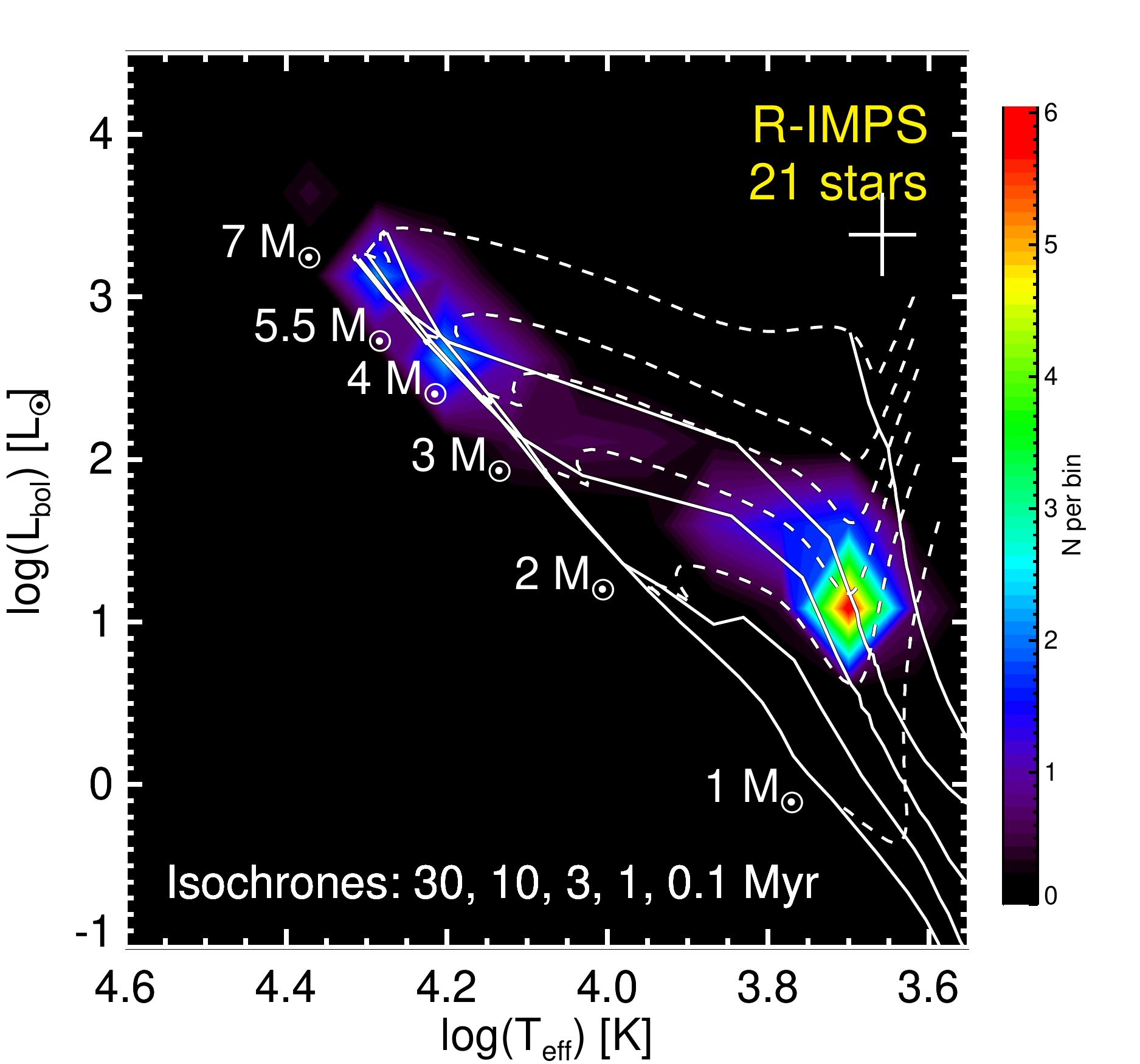}
    \includegraphics[scale=0.4]{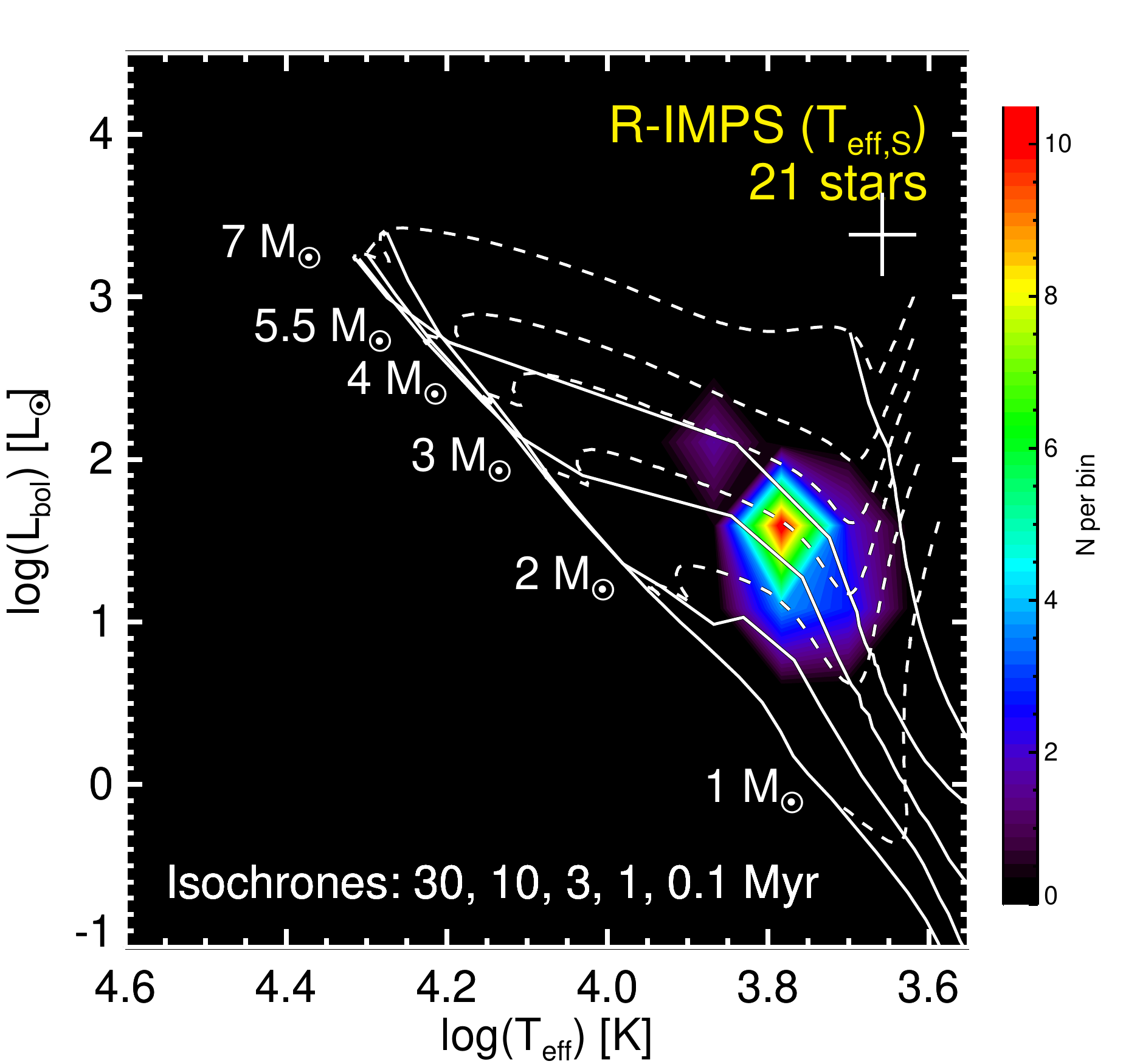}
    \caption{Probabilistic Hertzsprung-Russell Diagrams for SED fits to the 21 diskless sources with available $T_{\rm eff,S}$ from spectroscopy (D17). \textit{Left}: Results using our standard model parameter age-weighting functions (Section \ref{sec_src_class}). \textit{Right}: Results using the $T_{\rm eff,S}$ constraints (Section \ref{sec_ir_damiani}). Overlays and annotations are the same as in Figure~\ref{fig_phrd}. 
    \label{fig_rimps_phrd}
    }
\end{figure*}
 The two composite pHRDs plotted in Figure~\ref{fig_rimps_phrd} show why the $T_{\rm eff,S}$ constraints are required to classify R-IMPS. Our age-weighting scheme for SED models seldom preferred $T_{\rm eff}$ that was close to the actual $T_{\rm eff,S}$ values for R-IMPS. This is mainly due to the low density of SED models populating this range of $T_{\rm eff}$ and $L_{\rm bol}$ in both the naked PMS and \citet{robitaille_2006} model sets. Based on photometry and our age-weighting alone, the most likely models are typically cooler, and occasionally hotter (P19).  There are almost certainly additional R-IMPS in our sample that lacked spectroscopic measurements, in particular those located in the wider field outside the relatively small, central region surveyed by D17. In the absence of $T_{\rm eff,S}$ constraints, 
 about 60\% of R-IMPS were misclassified as cooler, less luminous IMPS, and the remainder were predominantly misclassified as hotter, more luminous AB stars.

\deleted{
\begin{table}[htb]
\centering
\caption{TM Classes} \label{tab_tm_classification}
    \begin{tabular}{ccc}
    \hline
    \hline
    Classification  &  $T_{\rm eff}$\tablenotemark{a} (K) &   $M_\star$ ($M_{\odot}$) \\
    \hline
    IMPS                &   ${\leq}7300$ &  2--8 \\
    R-IMPS              &   ${\geq}5300$\tablenotemark{b} &  2--8 \\
    TTS                 &   ${\leq}7300$ &  ${\leq}2$     \\
    AB                  &   $>$ 7300    &  2--8  \\
    OB                  &  \nodata   &   ${>}8$ \\
    \hline
    \end{tabular}
        \tablenotetext{a}{$T_{\rm eff}=7300$~K is the canonical value for an F0 V star.} 
        \tablenotemark{b}{R-IMPS require spectroscopic $T_{\rm eff,S}$}
\end{table}
}
\subsection{X-ray Spectral Fitting}\label{sec_xspec}
We fit the 0.5--8 keV X-ray spectra of all sources in our sample with \texttt{XSPEC} v12.91.1 \citep{arnaud_96} \edit1{adopting solar abundances from \citet{grevesse+1998}}. We model each source using either one thermal plasma (\texttt{apec}) component (which we refer to as the 1T model) or two \texttt{apec} components \citep[the 2T Model;][]{smith+2001}. The plasma temperature $kT$ in the 1T models is left as a free parameter. In the 2T models, one temperature is fixed at $kT_1=0.9$~keV (corresponding to the expected ${\sim}10$~MK base emission for coronal TTS emission observed by P05), while $kT_2$ is left as a free parameter.

Both the 1T and 2T models include an absorbing column $N_{\rm H}$ \citep[$\texttt{tbabs}$;][]{wilms+2000}, which can be left as a free parameter in the fits or \edit1{frozen to the equivalent absorption corresponding to the extinction  ($A_V$) value returned by the IR SED fits or NIR colors for the source.}
For sources fit with the \citet{robitaille_2006} YSO models, this extinction is the sum of the foreground $A_V$ plus the total $A_V$ through the circumstellar dust disk and/or infalling envelope.\footnote{In some cases the internal $A_V$ parameter for the disk+envelope can be very high, and it may be highly uncertain or even unreliable, in which case the Free models were preferred. The large majority of X-ray bright sources in our sample were fit with diskless PMS models that are  unaffected by this issue. The internal $A_V$ parameter is no longer implemented in the updated YSO model sets of \citet{robitaille17}.} $A_{\rm V}$ was converted to $N_{\rm H}$ using the relation $N_{\rm H}/A_V = 1.6\times 10^{21}$~cm$^{-2}$~mag$^{-1}$ \citep{vuong_2003}. We hence further divide our X-ray spectral fitting results into ``Free'' or ``\edit1{Frozen}'', for which $N_{\rm H}$ was left as a free parameter or constrained using the $A_V$ from IR SED fitting or NIR colors (when the SED fitting returned \edit1{unrealistically large or poorly-constrained} $A_V$ values). 
Table \ref{tab_xspec_models} summarizes the naming scheme that we use for the X-ray spectral models.

\begin{table}[htb]
\centering
\caption{\edit1{X-ray Model Naming Scheme}} \label{tab_xspec_models}
    \begin{tabular}{ccc}
    \hline
    \hline
                                    &1T Plasma  &2T Plasma  \\   
    \hline
    Free N$_H$                      &1T Free    &2T Free    \\  
    $A_{\rm V}$-Frozen N$_H$        &1T Frozen  &2T Frozen  \\

    \hline
    \end{tabular}
\end{table}

\begin{figure*}[htb]
    \centering
    \includegraphics[scale=0.118]{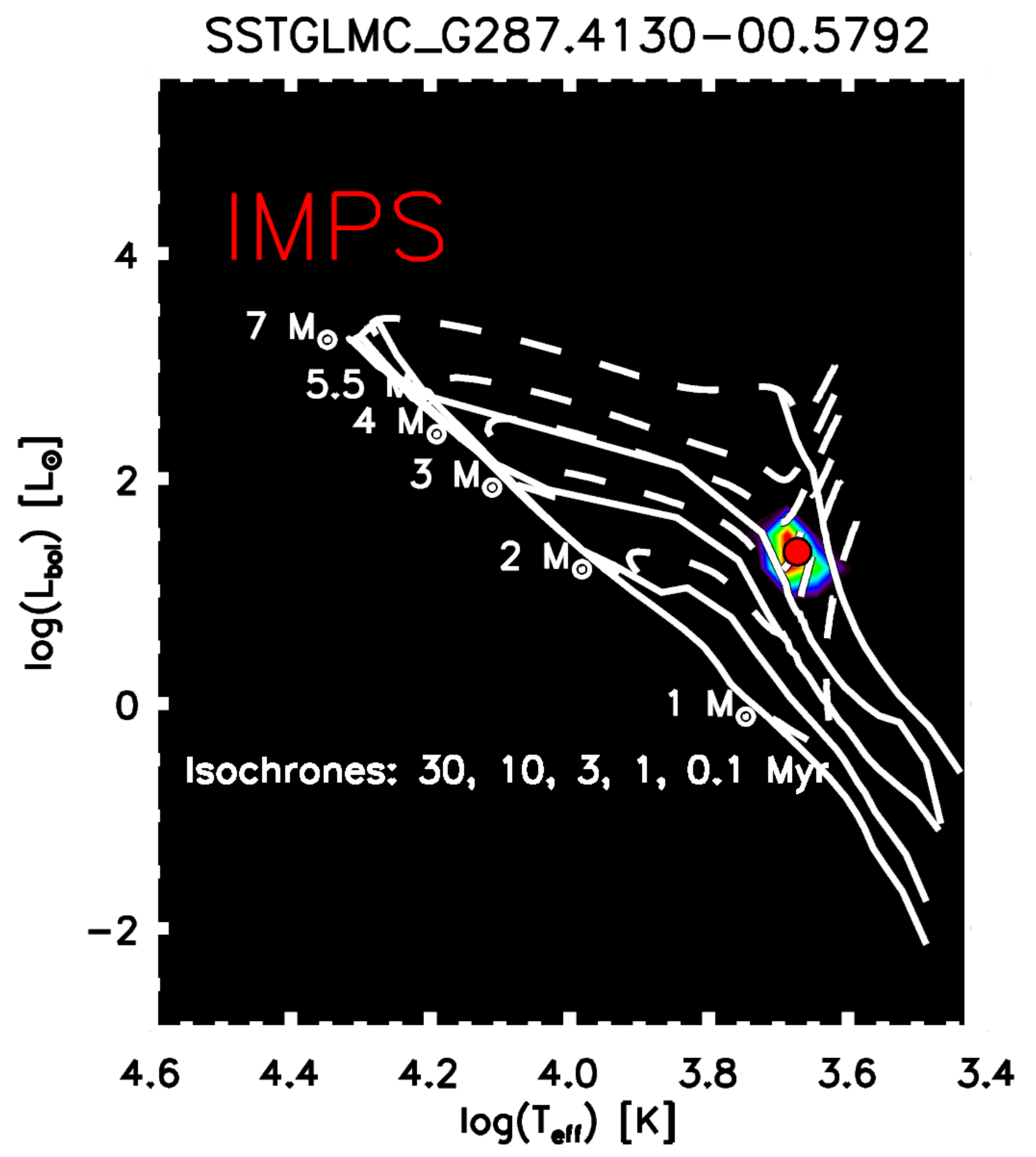}
    \includegraphics[scale=0.25]{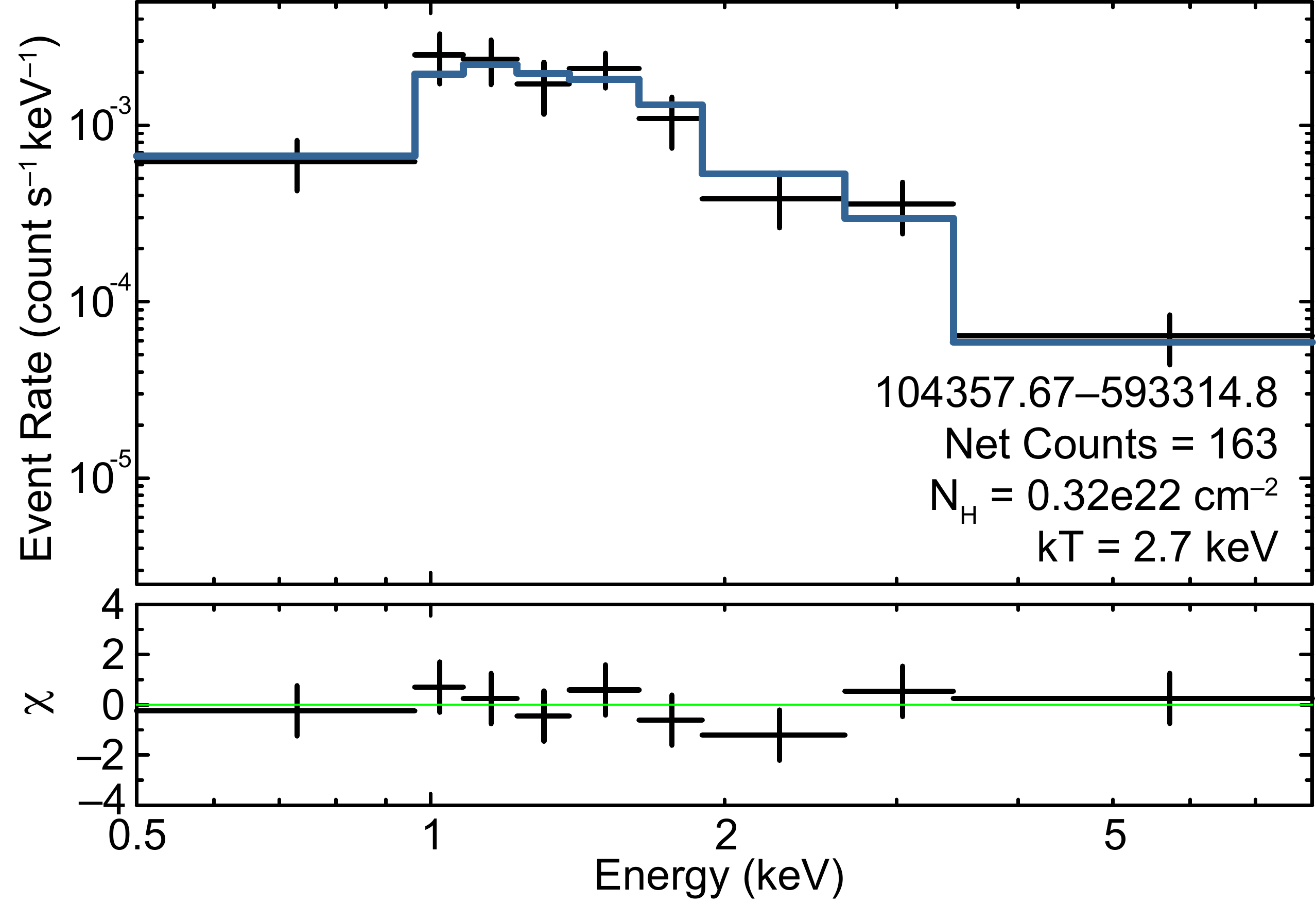}
    
    \includegraphics[scale=0.17]{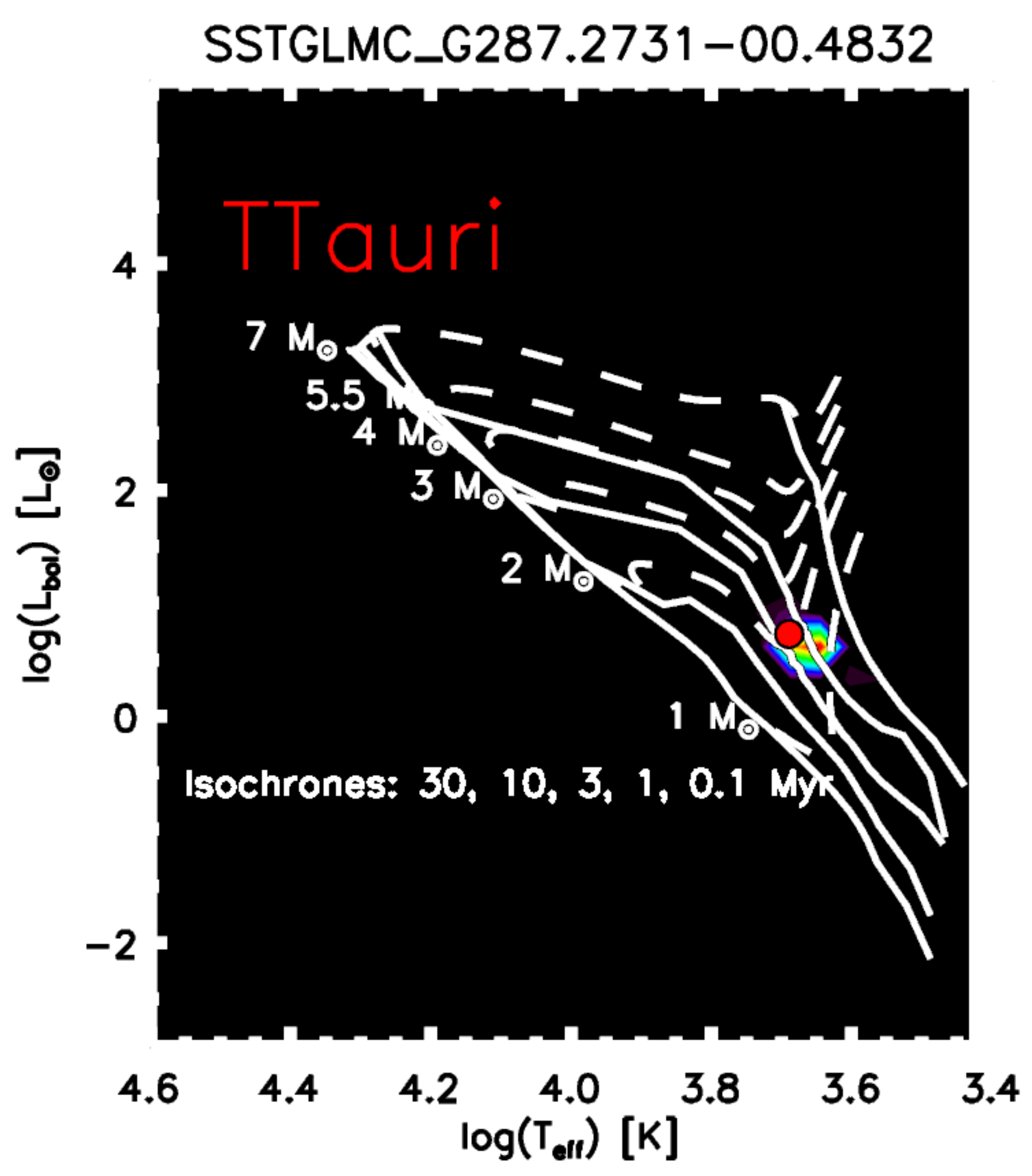}
    \includegraphics[scale=0.25]{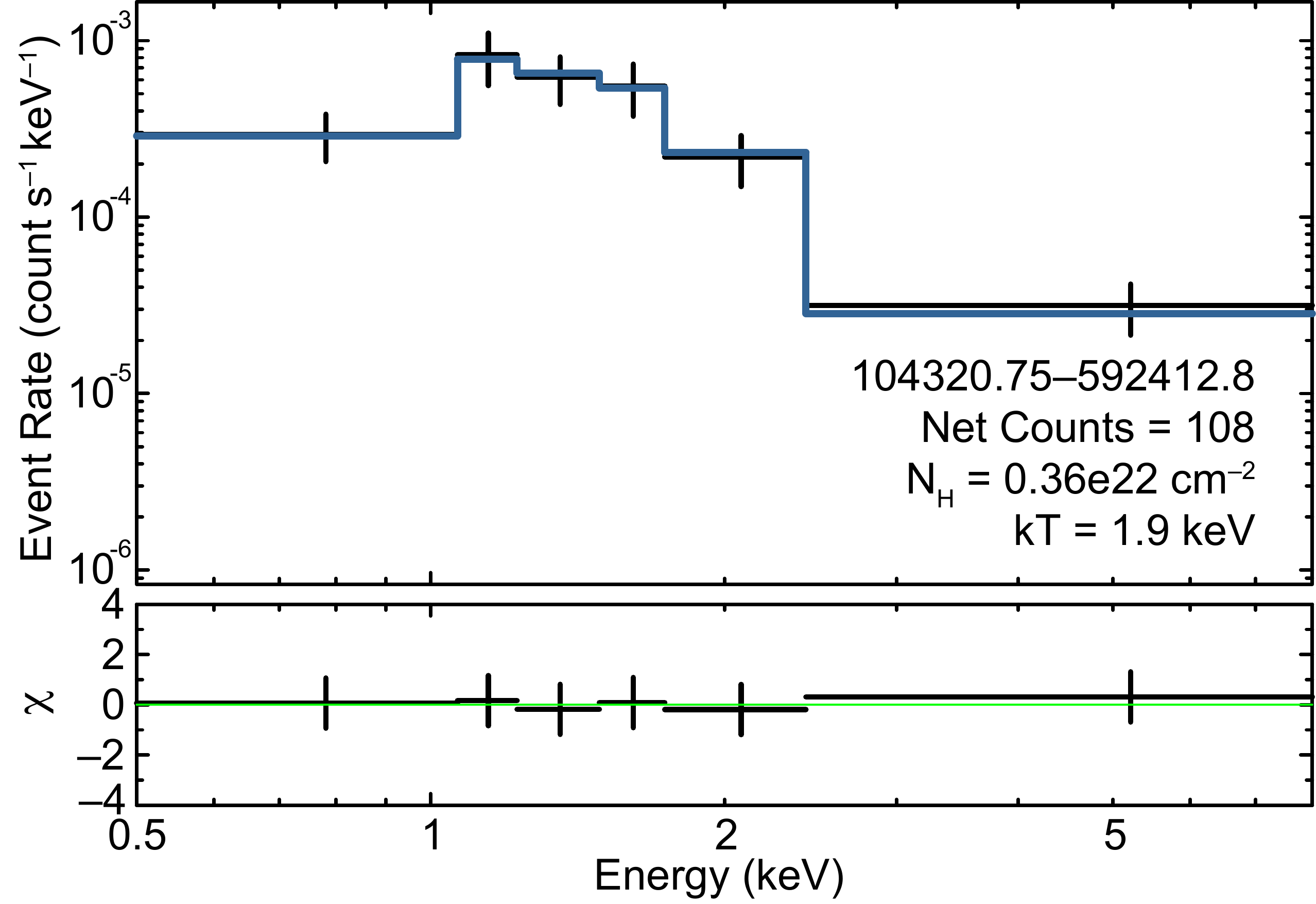}
    
    \includegraphics[scale=0.118]{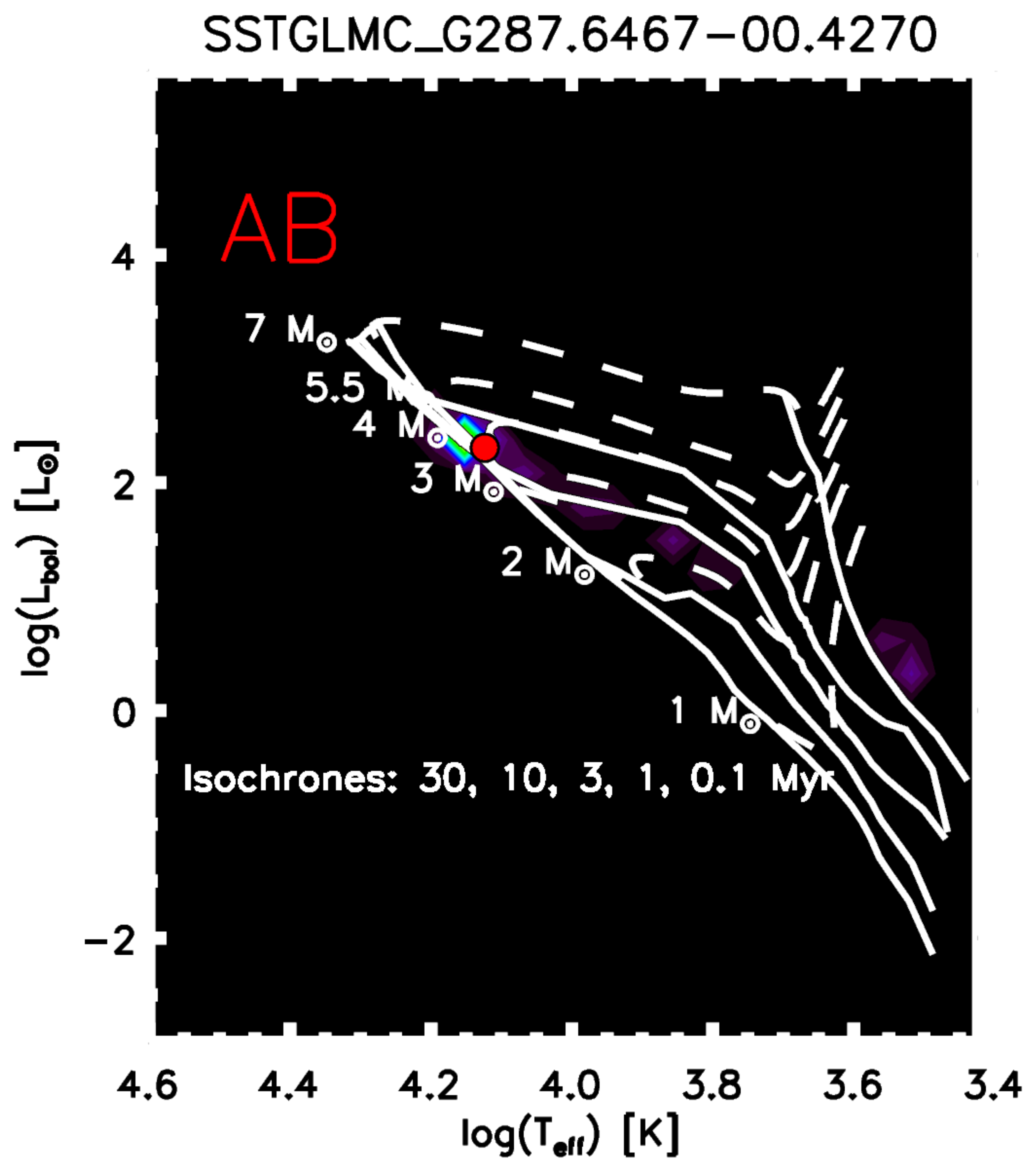}
    \includegraphics[scale=0.25]{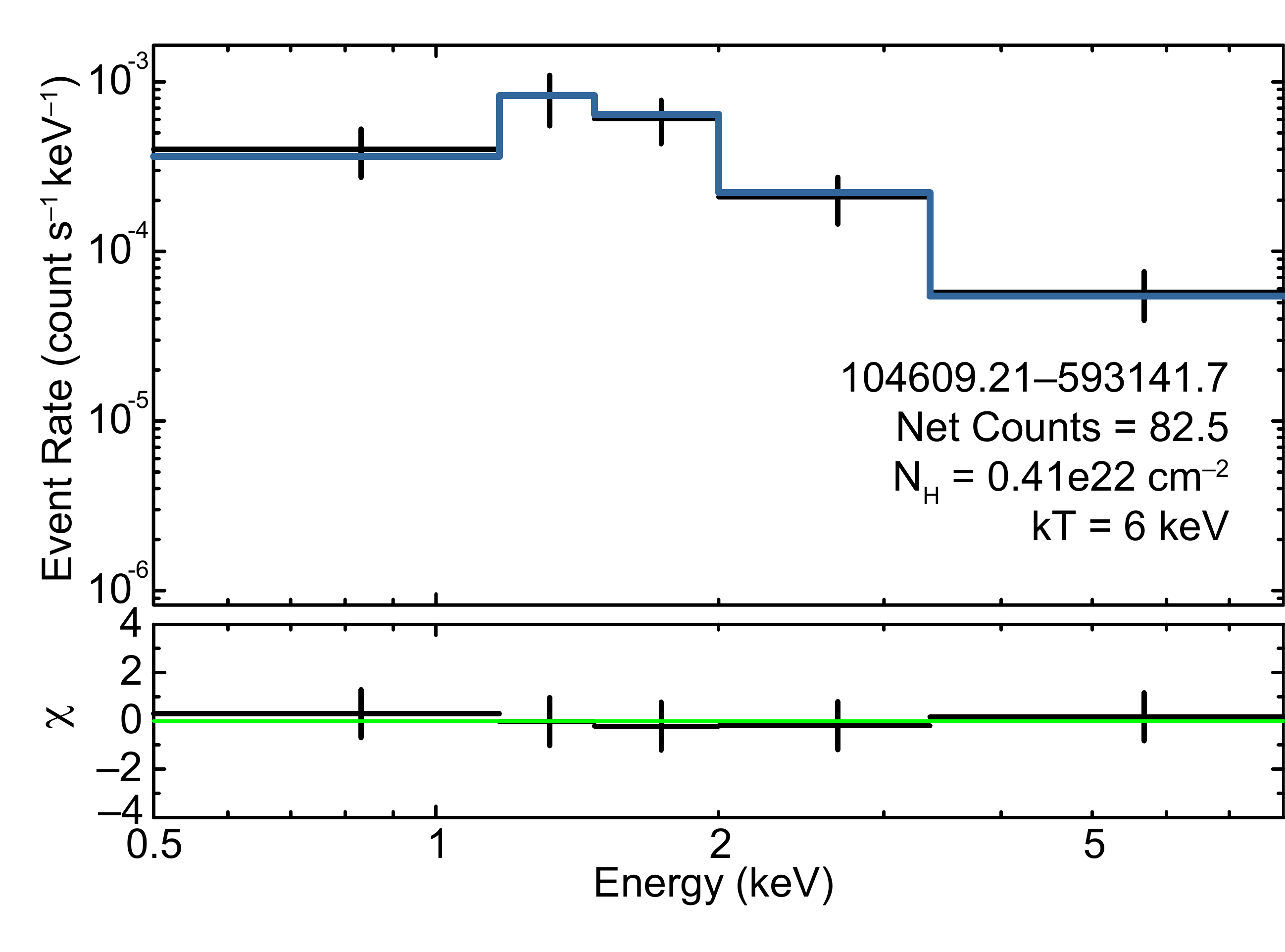}
    
    \includegraphics[scale=0.17]{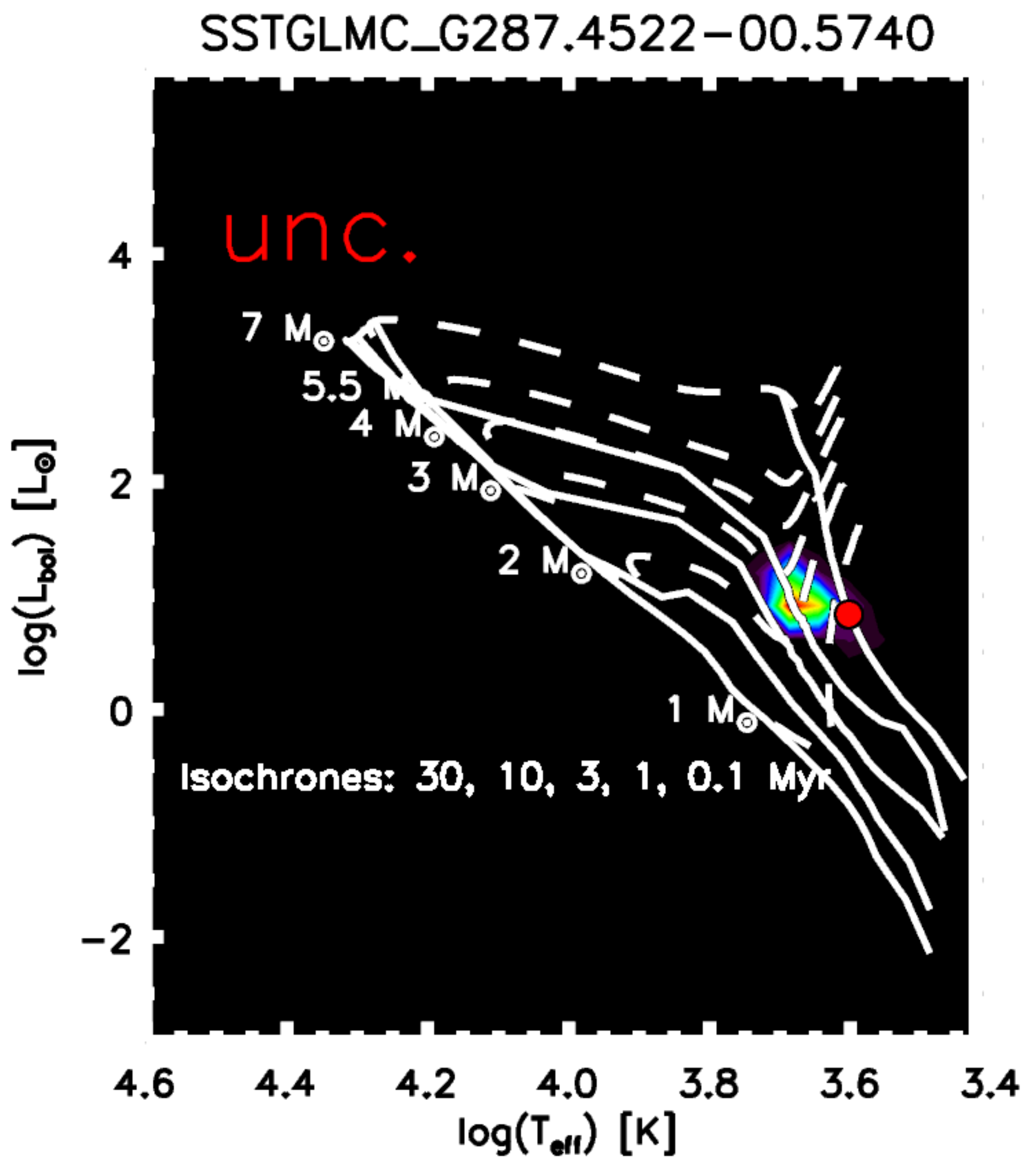}
    \includegraphics[scale=0.25]{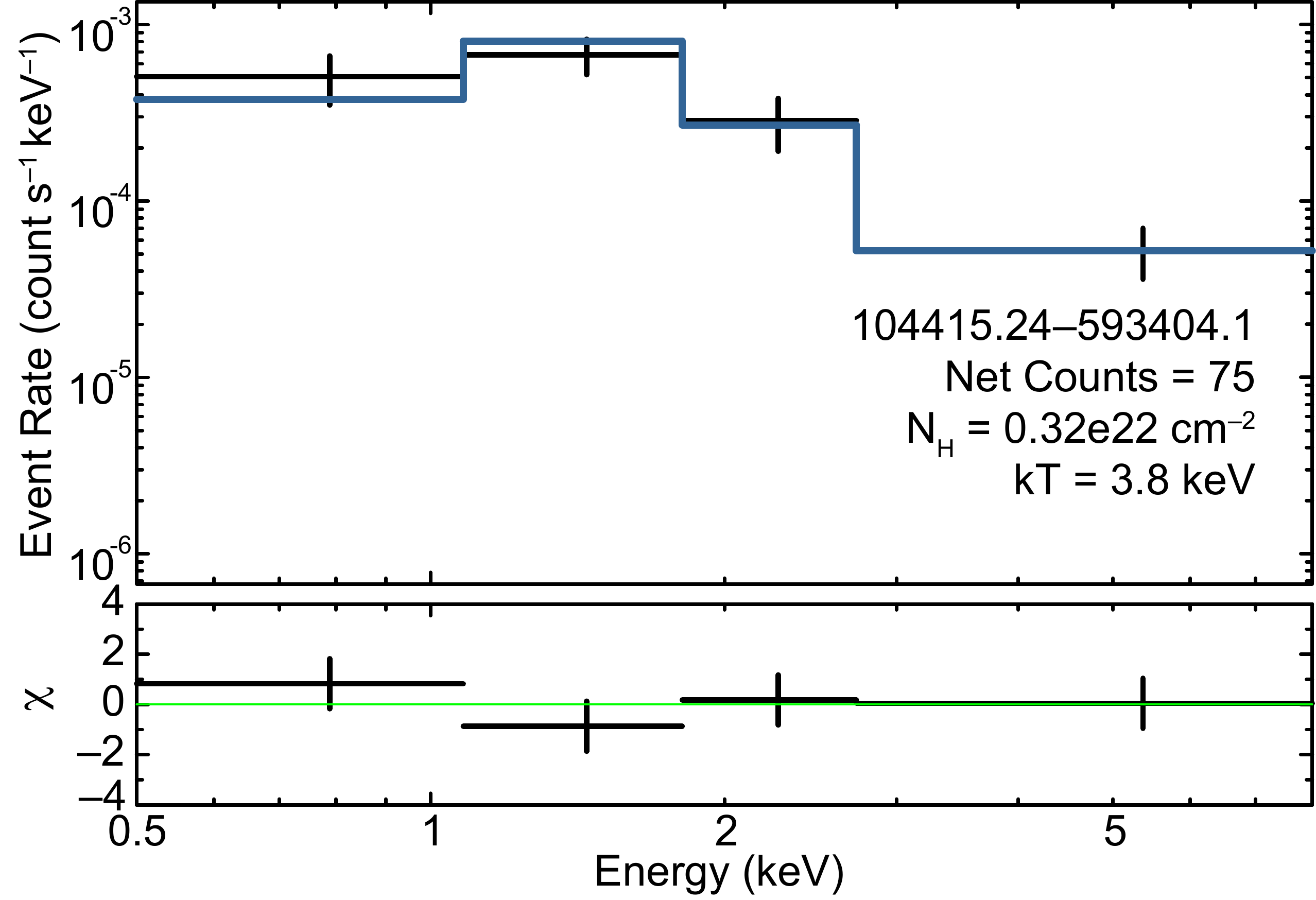}
    
    \caption{pHRDs (left column) and 0.5-8 keV X-ray spectral fits with residuals (black=data with error bars, blue=best-fit spectral model; binned for display purposes only) for example sources representing the various $TM$-classes.
    \label{fig_xspec_phrd}
    }
\end{figure*}

The models were fit to the unbinned X-ray spectra using the Cash statistic, $C$ \citep{Cash79}. 
    Figure \ref{fig_xspec_phrd} gives examples of both the X-ray spectrum, \edit1{binned for display purposes only,} and individual pHRD of a single source from each $TM$ Class (Section \ref{sec_src_class}). 
     \edit1{These visualizations of the X-ray spectral fits} and pHRDs for our full CCCP X-ray bright source catalog are available for download from our \href{https://doi.org/10.5281/zenodo.4628273}{Zenodo repository}.\footnote{DOI:10.5281/zenodo.4628273} 

We visually reviewed all \texttt{XSPEC} model fits to each of our X-ray spectra using the GUI-based tool \linebreak {\scshape ae\_spectra\_viewer} provided with {\scshape ACIS Extract} \citep{broos_2010}. In the majority of cases, the fit statistics (e.g., the Cash statistic $C$ per degrees of freedom) between the four models were similar.  We therefore chose the best-fit spectral model among the 1T Frozen, 2T Frozen, 1T Free, and 2T Free based upon the residuals and grouped spectra outputs (shown in the right column of Figure \ref{fig_xspec_phrd}).

\edit1{When all models showed similar fit quality we preferred the 1T Free model for simplicity.} 
\edit1{Typical uncertainties seen in $kT$, when upper and lower limits are found (${\sim}2/3$ of our sample), have a range of ${\pm}2~\rm keV$. The other 1/3 of the fits returned only lower limits on $kT$. About half of our sample fits returned doubly-bounded uncertainties on $N_{\rm H}$, typically ranging over ${\pm}0.5\times10^{22}~{\rm cm^{-2}}$. The other half of the sample returned only upper limits on $N_{\rm H}$. 
There are some cases ($\sim10\%$ of our sample) where $kT$ or $N_H$ do not have reported errors; in these cases the parameters were at the edge of the allowed parameter space (i.e. $kT \;=\; 10\;keV$ or $N_{\rm H}$ = (0.09--33)$\times 10^{-22} \; \rm{cm^{-2}}$). There are no sources that are missing both $N_{\rm H}$ and $kT$ errors.
There were a small number of cases (8 total) where the Free models did not fit the spectra well and we froze $N_{\rm H}$ to the absorbing column converted from the weighted mean $A_{\rm V}$ of the IR source, which achieved good fits and reported $kT$ uncertainties.}


 We removed 32 sources that could not be adequately fit with any of the models included in Table \ref{tab_xspec_models}.\footnote{These removed sources were not included in the composite pHRDs of Figures~\ref{fig_phrd} and \ref{fig_rimps_phrd}.} The majority of these had high X-ray background levels (${\ga}30\%$ of source counts) causing poor modeling of both the source and the background. \edit1{The distribution of $TM$ classes among our final sample is given in the ``Final'' column of Table \ref{tab_src_cnt}.}

\begin{figure*}[htp]
    \centering
    \includegraphics[scale=0.96]{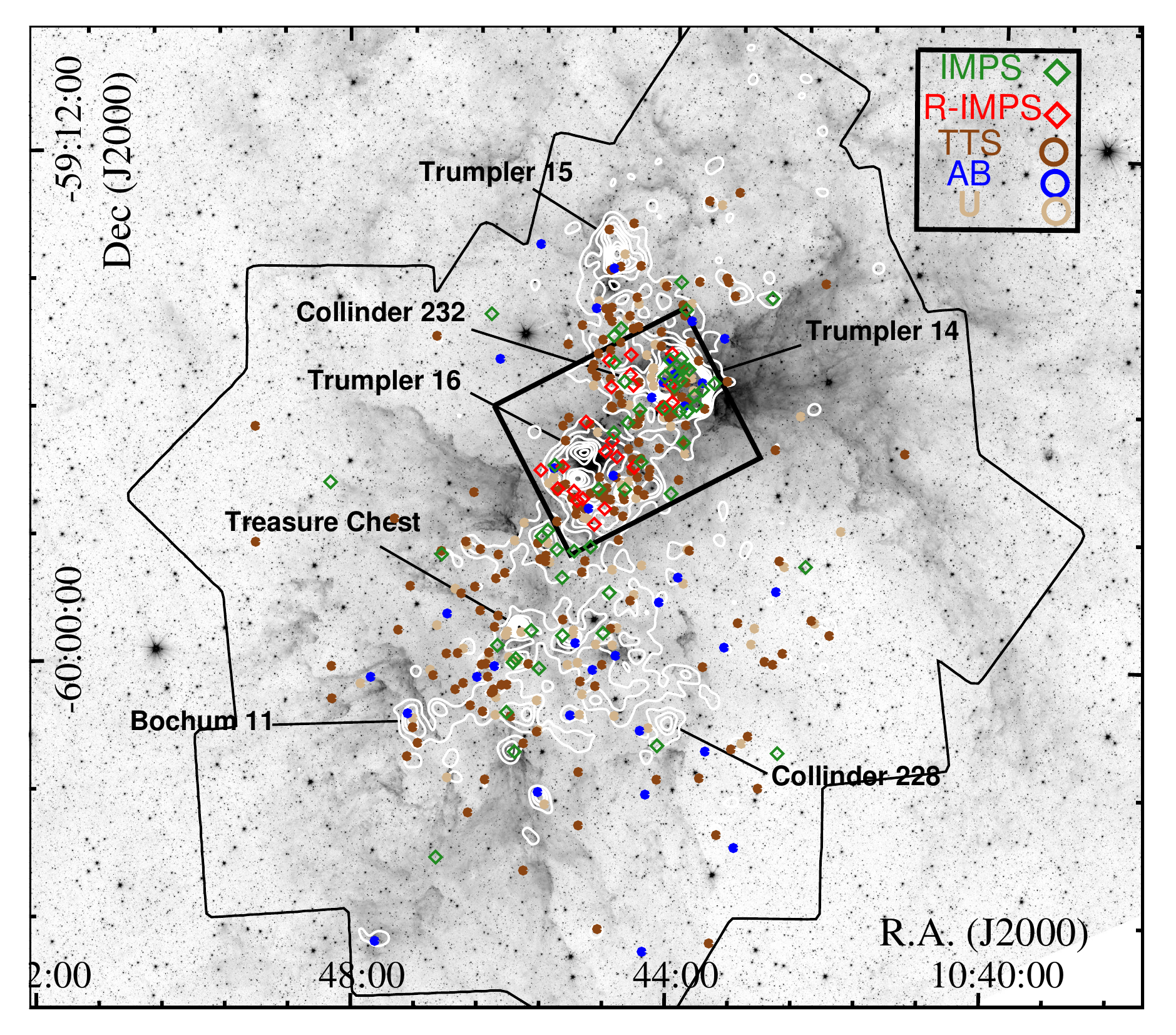}
    \caption{\textit{Spitzer}/IRAC $3.6 \; \mu$m mosaic image (inverted logarithmic grayscale) of the Carina Nebula. The 54 IMPS candidates are marked by green diamonds; 23 R-IMPS by red diamonds; 176 TTS by brown circles; 35 AB by blue circles; and 82 U by tan squares. 
    The thick black box in the center of the field outlines the D17 survey area. The white contours show X-ray point source density from \citet{feigelson_2011}; Tr 14 contains the highest concentration of IMPS. The thin black outline shows the edges of the CCCP X-ray survey area \citep{townsley_2011}.
    \label{fig_carina}
    }  
\end{figure*}

The sky positions of all 370 sources in our final sample are overlaid on a \textit{Spitzer} 3.6~\um\ mosaic of the Carina Nebula in Figure \ref{fig_carina}. The colored symbols denote stars in their respective $TM$ Classes, while the white contours correspond to the source density of X-ray young stellar members \citep{broos_2011,feigelson_2011}. Every $TM$ Class is represented in all parts of the region. IMPS and TTS show more clustering compared to AB and U sources. IMPS are densely clustered in Tr 14, which contains roughly one-third of all IMPS in our sample and has the highest X-ray point source density of the CCCP catalog \citep{feigelson_2011}. Because R-IMPS require spectroscopic confirmation for classification, all are found within the Gaia-ESO survey area (black thick box; D17). The majority of the R-IMPS (14/23) cluster around Tr 16 (\citet{wolk+2011}), which contains the famous massive binary system $\eta$ Carinae \citep{hamaguchi+2014}. 
X-ray bright TTS generally follow the spatial distribution of all X-ray detected members (contours), especially in the vicinity of Tr 16 and  Tr 14. 

\section{X-ray Spectral Analysis Results} \label{sec_analysis}
\subsection{Best Fit Spectral Models} \label{sec_best_model}
In Table \ref{tab_best_fit} we tally for each $TM$ Class the best fit \texttt{XSPEC} model type (1T or 2T),  origin of the $N_{\rm H}$ parameter (free in the fit or \edit1{frozen} by $A_V$) and occurrence of variability in the X-ray light curve. 
We use the same variability criterion described in detail by \citet{broos_2010}, by which the light curves of each source during a single observation is tested for uniform flux over time using a one-sided Kolmogorov-Smirnov (KS) Test. P-values ($Prob_{KS}$) for the null hypothesis of uniform flux are flagged as ``no variability'' (0) for \texttt{$Prob_{KS}$} $> 5 \times 10^{-2}$, ``possible variability'' (1) for $5 \times 10^{-2} <$ \texttt{$Prob_{KS}$} $< 5 \times 10^{-3}$, and ``definite variability'' (2) for \texttt{$Prob_{KS}$} $< 5 \times 10^{-3}$.  
Although a multi-observation KS probability is computed by {\scshape ACIS Extract} it is not used for determining variability in this work. 
The single-observation exposure times vary across the large CCCP mosaic, but are generally ${<}60$~ks, a timescale well-matched to the duration of individual coronal flares \citep{favata+2005}.

\begin{table}[h!]
\begin{center}
\caption{Best Fit Source Statistics} \label{tab_best_fit}
    \begin{tabular}{cccccccc}
    \hline
    \hline

  -                         &1T             &2T             &Frozen         &Free           &1T \&          &2T \&      &\% \\
                            &               &               &               &               &Vary           &Vary       &Vary\\ 
    \hline
    IMPS                    &\edit1{53}     &\edit1{1}      &\edit1{0}      &\edit1{54}     &10             &1          &20\\
    R-IMPS                  &23             &0              &\edit1{0}      &\edit1{23}     &4              &0          &17\\
    TTS                     &\edit1{176}    &\edit1{0}      &\edit1{6}     &\edit1{170}    &\edit1{60}     &\edit1{0}  &34\\
    AB                      &34             &1              &\edit1{0}      &\edit1{35}     &7              &1          &23\\
    U                       &\edit1{82}     &\edit1{0}      &\edit1{2}     &\edit1{80}     &\edit1{20}     &\edit1{0}  &24\\
    All\tablenotemark{b}    &\edit1{368}    &\edit1{2}      &\edit1{8}     &\edit1{362}    &\edit1{101}    &\edit1{2}  &28\\
    
    \hline
    \end{tabular}
    \end{center}
        \tablenotetext{a}{Sum of all rows in a single column. The total number of sources per population = 1T + 2T or \edit1{Frozen} + Free. There are 370 total sources.}
\end{table}

We found that for $\sim$82\% (306/370 sources) of our sample, the $N_{\rm H}$ predicted from the best-fit Free models (including cases where the Frozen model was preferred) fell within the range of $A_V$ returned by the IR SED fitting or estimated from NIR photometry. \edit1{This high frequency of
agreement between the absorption inferred via our independent X-ray and IR modeling analysis increases our confidence in the physical parameters output by both models. The degeneracy between absorbing column and temperature (plasma or photospheric) is the greatest challenge to our modeling in both cases.}

The overwhelming majority of sources in our sample (\edit1{99\%; Table~\ref{tab_best_fit}}) required only a single plasma component (1T model) to achieve acceptable \texttt{XSPEC} fits. Even among our sample of bright CCCP X-ray sources, the majority were detected with ${<}100$ net counts. But in cases where $L_X$ is relatively high and $N_H$ is non-negligible, a single (harder) component can dominate the observed spectrum\edit1{; the high $N_H$ makes it difficult to observe the soft emission since it is more readily absorbed. This leads to a lack of information about the contribution from lower coronal temperatures.}

Using the best-fit model $N_{\rm H}$ to correct observed fluxes for absorption, we computed the hard-band ($L_{h,c}$; 2--8~keV) and total-band ($L_{t,c}$; 0.5--8~keV) X-ray luminosities assuming the Gaia DR2 parallax distance to Carina of 2.5~kpc (P19). We prefer hard-band luminosities over over soft-band (0.5--2~keV) luminosities for this analysis because the hard photons are less sensitive to absorption correction. 
\edit1{A recent distance determination based on Gaia EDR3 to a subset of massive stars in Carina reports 2.3~kpc \citep{shull+2021}. The discrepancy with our adopted distance, which is based on a much larger sample of X-ray detected stars, implies there may be a 15\% systematic uncertainty in our reported luminosities.}

The best fit X-ray and IR properties for our sample will be available as an electronic table. The columns of the e-table are described in Table \ref{tab_list_parameters}.


\begin{table*}[htbp]
\centering
\scriptsize
\caption{X-ray and IR Source Properties} \label{tab_list_parameters}
    \begin{tabular}{lcl}
    \hline
    \hline
    Column Label            &Units                  &Description    \\
    \hline
    XName                   &\nodata                &IAU source; prefix is CXOGNC J(\textit{Chandra X-ray}) 
                                                    \textit{Observatory Great Nebula in Carina} \\
    MIRName                 &\nodata                &\textit{Spitzer} MIR Vela-Carina source name  \\
    RA                      &deg                    &Right Ascension (J2000)    \\
    Dec                     &deg                    &Declination (J2000)    \\
    TMClass                 &\nodata                &$Temperature-Mass$ Classification 
                                                    (Section \ref{sec_src_class})    \\
    YSO                     &\nodata                &(y)es or (n)o; Source a YSO 
                                                    (See Section \ref{sec_observations}) \\
    \edit1{FrozenModel}        &\nodata              &\edit1{(y)es or (n)o; $N_{\rm H}$ frozen to mean IR $A_{V}$ 
                                                    (See Section \ref{sec_xspec})}  \\
    \edit1{CStat\_DOF}              &\nodata                &\edit1{C-Statistic/Degrees Of Freedom for X-ray spectral fit}    \\
    NetCounts               &count                  &Full band (0.5 - 8 keV) net counts     \\
    Variable                &\nodata                &(0) non-variable, (1) possible, or (2) definite; 
                                                    (See Section \ref{sec_best_model}) \\
    SrcArea                 &(0.492arsec)$^2$       &Average aperture area  \\
    Theta                   &arcmin                 &\textit{Chandra} off axis angle  \\
    Mass                    &$M_\odot$              &Mass from SED  \\
    Mass\_err               &$M_\odot$              &Weighted 1--sigma uncertainty of $\log$Mass (Section \ref{sec_src_class}, P19)  \\   
    NHX                     &10$^{22}$ cm$^{-2}$    &Hydrogen absorbing column from XSPEC   \\
    \edit1{NHX\_err\_lo}     &\edit1{10$^{22}$ cm$^{-2}$}    &\edit1{90\% confidence interval lower bound on NH\_X}   \\
    \edit1{NHX\_err\_hi}     &\edit1{10$^{22}$ cm$^{-2}$}    &\edit1{90\% confidence interval upper bound on NH\_X} \\
    AvSED           &\nodata                &(y)es or (n)o; See Section \ref{sec_best_model} \\
    Av\_mean        &mag                    &Mean on Av; When AvSED = y, 
                                                    Av from SED fit displayed. Else, Av from NIR estimate is 
                                                    displayed   \\
    Av\_min         &mag                    &Lower bound of weighted 1--sigma uncertainty on Av (Section \ref{sec_src_class}, P19)   \\
    Av\_max         &mag                    &Upper bound of weighted 1--sigma uncertainty on Av (Section \ref{sec_src_class}, P19)   \\
    2TModel                 &\nodata                &(y)es or (n)o; Source fit with 2T plasma component   \\
    kT                      &keV                    &1T or 2T plasma temperature; When ``2TModel'' = y, 
                                                    kT2 is displayed (See Section \ref{sec_xspec}) \\
    \edit1{kT\_err\_lo}     &\edit1{keV}                    &\edit1{90\% confidence interval lower bound on kT}    \\
    \edit1{kT\_err\_hi}     &\edit1{keV}                    &\edit1{90\% confidence interval upper bound on kT}    \\
    Fx\_hc                  &erg s$^{-1}$ cm$^{-2}$ &Hard band (2--8 keV) absorption corrected flux    \\
    Fx\_tc                  &erg s$^{-1}$ cm$^{-2}$ &Total band (0.5--8 keV) absorption corrected flux    \\
    logLx\_hc               &(erg s$^{-1}$)         &Hard band (2--8 keV) absorption corrected luminosity  \\
    logLx\_tc               &(erg s$^{-1}$)         &Total band (0.5--8 keV) absorption corrected luminosity  \\
    logLbol                 &(L$_\odot$)            &Bolometric Luminosity from IR SED  \\
    logLbol\_err            &(L$_\odot$)            &Weighted 1--sigma uncertainty on $\log$Lbol (Section \ref{sec_src_class}, P19)  \\    
    
    \hline
    \end{tabular}
\end{table*}

\subsection{Absorption and Plasma Temperature Parameter Distributions}
We compared the distributions of $N_{\rm H}$ and $kT$ for each $TM$ Class against one another using two-sided KS tests. 
We define a ``significant difference'' as KS Probability $\lesssim 10^{-4}$, marginal as $0.01-10^{-4}$, and no difference as $> 0.01$. 
These KS tests reveal no statistically significant differences in $N_{\rm H}$ distributions among the $TM$ classes. 
In terms of average absorption within each $TM$ class, the AB sources have the lowest, at $(4\pm 2)\times10^{21} \; \rm{cm}^{-2}$, while TTS have the highest absorbing column, and largest spread, with values $<(40)\times10^{21} \; \rm{cm}^{-2}$. 
Similarly we find no statistically significant differences in plasma temperatures between the $TM$ Classes. We performed the same KS tests on the $kT1$ of the sources that were fit with 1T models (both Free and Frozen). The average $kT1$ for each $TM$ Class fell within the narrow range of 2.6--3.2 keV, corresponding to coronal plasma temperatures of 30--37 MK. 
Since there were very few sources that preferred the 2T models \edit1{(2/370)} (Table~\ref{tab_best_fit}), statistical comparisons of the $kT2$ parameter were not performed. 

\deleted{We comment briefly that the average $kT2$ varied more significantly across the sub classes compared to the $kT1$ parameter with a spread of 4.6 - 9.5 keV (53.3 - 110 MK).} 

\subsection{X-ray Luminosity Functions} \label{sec_lum}

\begin{figure*}[ht]
    \centering
    \includegraphics[scale=0.39]{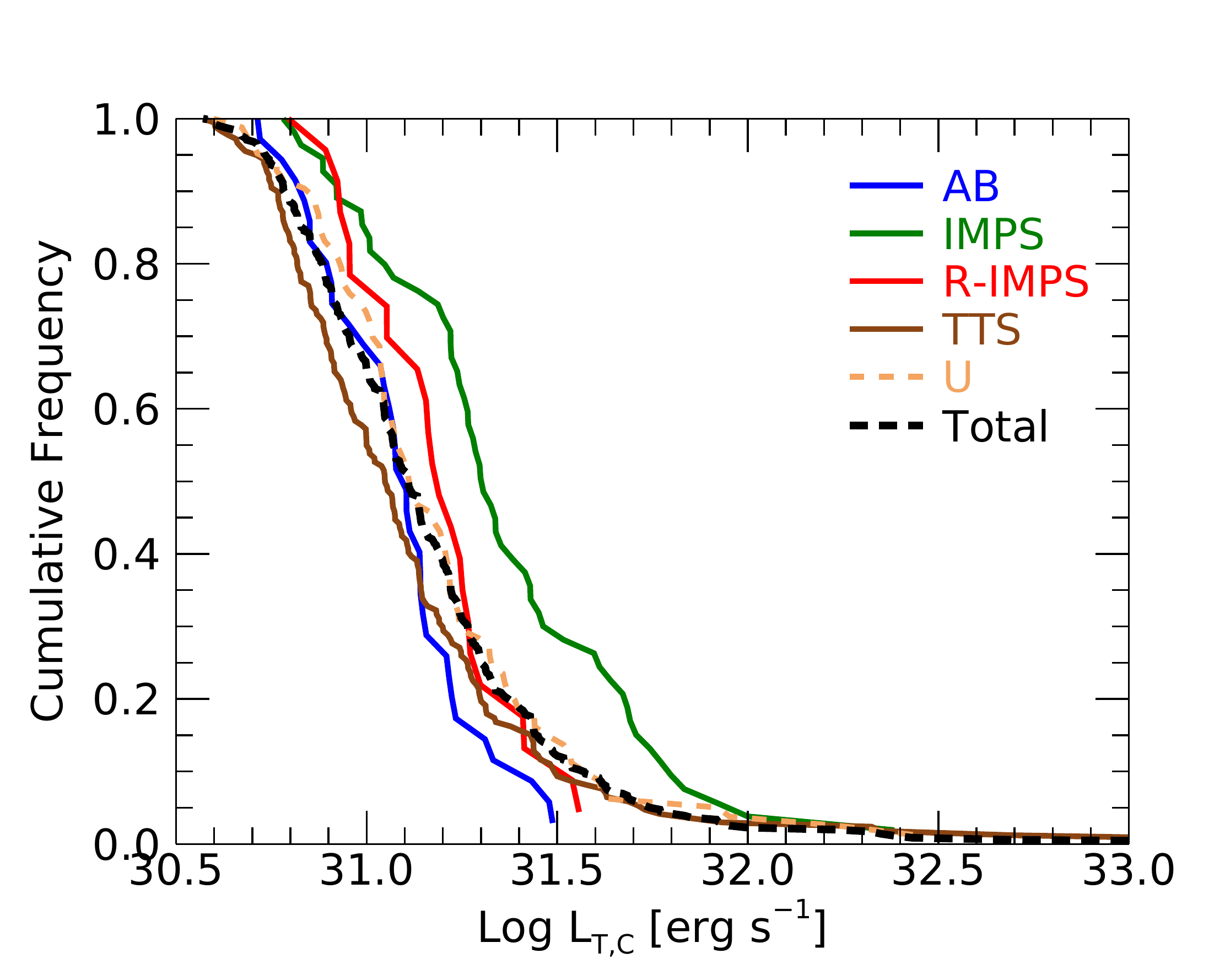}
    \includegraphics[scale=0.39]{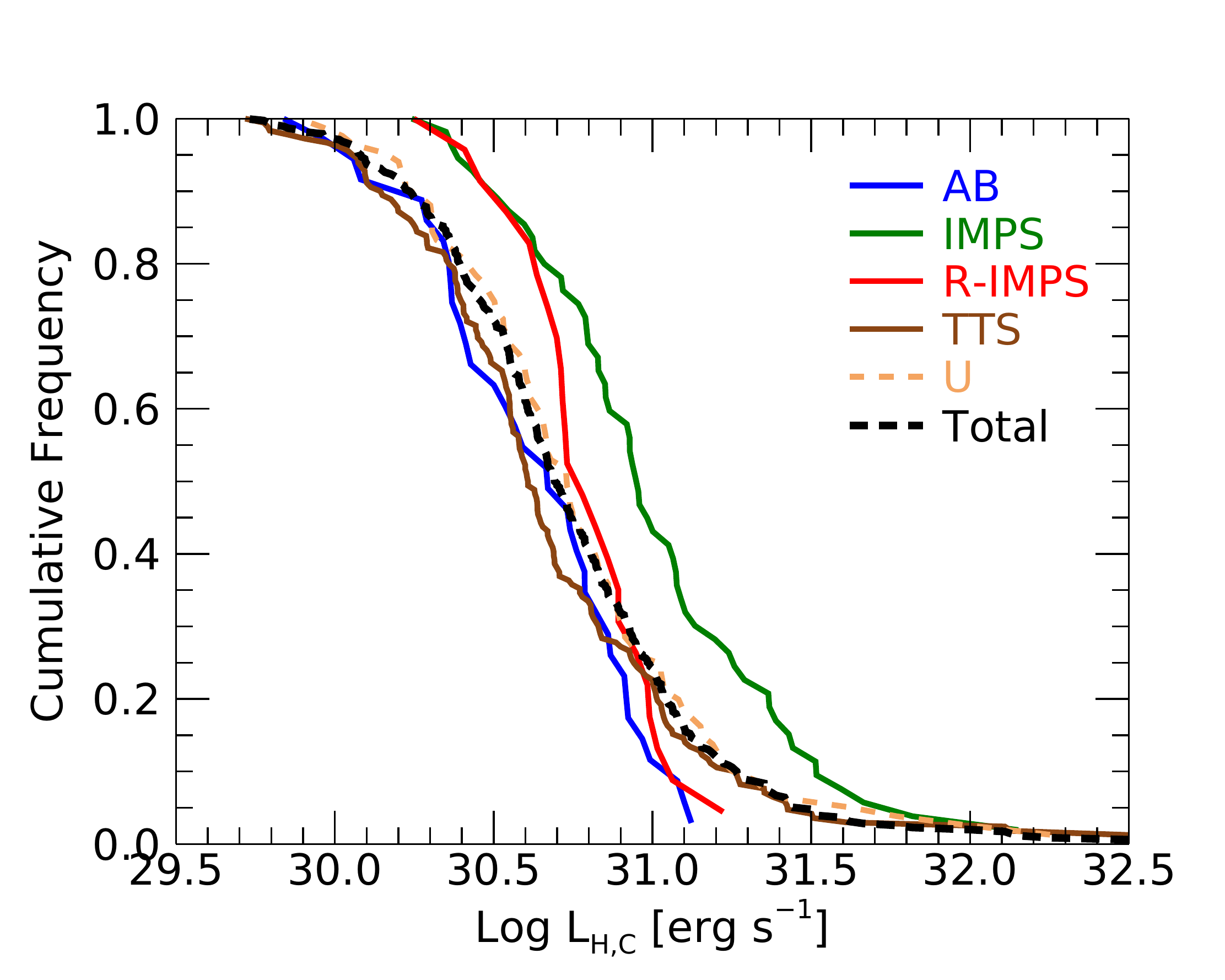}
    \caption{Absorption-corrected cumulative XLFs for the various $TM$ classes, shown for both 
     $L_{t,c}$ (0.5--8~keV; \textit{left panel}) and $L_{h,c}$ (2-8~keV; \textit{right panel}). IMPS are clearly more luminous than the other $TM$ classes. 
     In both panels, the faint limits of the R-IMPS and IMPS XLFs overlap, while the bright tails of the R-IMPS resemble the AB stars.
    \label{fig_lum_cuml}
    }
\end{figure*}

The cumulative X-ray luminosity functions (XLFs) for all $TM$ classes are shown in Figure \ref{fig_lum_cuml}. IMPS are systematically more luminous 
than the TTS, AB, and U classes,
with a mean offset of ${\sim}0.3$~dex in both $L_{t,c}$ and $L_{h,c}$. 
The most luminous R-IMPS (section \ref{sec_ir_damiani}), have comparable $L_{h,c}$ (Figure~\ref{fig_lum_cuml}, right panel) to the most luminous AB sources, while the least luminous R-IMPS are comparable to the least luminous IMPS. The resultant shape of the (partially-convective) R-IMPS XLF looks like a transition between that of the (fully-convective) IMPS and the (radiative) AB classes.
%
%
Most of the same qualitative trends among the $TM$ classes appear in the $L_{t,c}$ XLFs (Figure~\ref{fig_lum_cuml}, left panel), but the curves appear noisier because the absorption correction more strongly affects the soft (${<}2$~keV) portion of the spectrum.

Tables \ref{tab_lum_hard} and \ref{tab_lum_total} show the two-sided KS tests comparing $L_{h,c}$ and $L_{t,c}$ distributions for each class respectively. For the hard band luminosities (Table \ref{tab_lum_hard}) we find a statistically significant difference between IMPS and TTS. We also find marginally significant differences comparing IMPS with AB or U. KS tests reveal no statistically significant differences among the non-IMPS $TM$ classes. 
 KS tests applied to the total band luminosities (Table \ref{tab_lum_total}) give similar results, except for the inclusion of a statistically significant difference comparing IMPS with AB stars. 

\begin{table}[htb]
\centering
\caption{\edit1{KS Tests: Hard Band X-ray Luminosity}} \label{tab_lum_hard}
    \begin{tabular}{cccccc}
    \hline
    \hline
    -       &IMPS  &R-IMPS   &TTS   &AB   &Unc   \\
IMPS    &1   &    0.0319   &  \textbf{5.03e-7}   &  9.12e-4   &   0.0018
   \\
R-IMPS  &    0.0319 &1   &    0.0121   &     0.107   &     0.338
   \\
TTS     &  \textbf{5.03e-7}   &    0.0121   &1   &     0.621   &     0.126
   \\
AB      &  9.12e-4   &     0.107   &     0.621   &1   &     0.531
 \\
Unc     &   0.0018   &     0.338   &     0.126   &     0.531
   &1   \\
    \hline
    \end{tabular}
\end{table}


The XLFs have not been corrected for survey completeness. However, in the hard-band XLF (right panel of Figure \ref{fig_lum_cuml}) the faint-end turn-over for the AB/TTs/U curves occurs $\sim$0.5 dex fainter compared to IMPS. The faint end cutoff for the AB/TTs/U is likely due to incompleteness, but the IMPS cut-off seems to be real as it occurs at a luminosity $\sim0.5$ dex higher than the likely completeness cutoff at $\log{L_{h,c}} \sim 30.4$. This seems to indicate a lower bound of  $\log{L_{h,c}}\approx 30.5$~erg~s$^{-1}$ for the X-ray luminosity of convective IMPS.

\begin{table}[htb]
\centering
\caption{\edit1{KS Tests: Total Band X-ray Luminosity}} \label{tab_lum_total}
    \begin{tabular}{cccccc}
    \hline
    \hline
    -      &IMPS   &R-IMPS   &TTS   &AB   &Unc   \\
IMPS    &1   &    0.0379   &  \textbf{1.78e-7}   &  \textbf{1.96e-5}   &  1.38e-4
   \\
R-IMPS  &    0.0379   &1   &    0.0594   &    0.0871   &     0.500
   \\
TTS     &  \textbf{1.78e-7}   &    0.0594   &1   &     0.582   &    0.0472
   \\
AB      &  \textbf{1.96e-5}   &    0.0871   &     0.582   &1   &     0.282
   \\
Unc     &  1.38e-4   &     0.500   &    0.0472   &     0.282
   &1   \\
    \hline
    \end{tabular}
\end{table}

\section{Discussion} \label{sec_discussion}
\subsection{IMPS as Luminous, Coronal X-ray Emitters} 
We have found that the X-ray spectra from our X-ray bright sample of low- and intermediate-mass stars in the CCCP survey can generally be fit using thermal plasma models returning similar plasma temperatures, implying a similar underlying physical emission mechanism.
The median hard- and total-band absorption-corrected X-ray luminosities for IMPS are ${\sim}0.3$~dex more luminous compared to TTS and AB stars, and the XLF turnover for IMPS occurs at higher luminosity (Figure~\ref{fig_lum_cuml}). 
Because the IMPS sample is dominated by fully-convective stars, this strongly suggests a coronal origin for IMPS X-ray emission. As a class, the IMPS sample is also more luminous than our TTS sample (which is itself relatively X-ray bright); thus IMPS X-ray emission must be intrinsic, not attributable to unresolved, lower-mass companions.

 R-IMPS are the second-most luminous $TM$ class, and the XLF turnover coincides with that of IMPS in both the hard and total-band XLFs (Figure~\ref{fig_lum_cuml}). 
 This reflects the decay in X-ray luminosity from the initially high values exhibited by IMPS, as the convective dynamo gives way to the development of an interior radiative zone \citep{mayne+07,mayne_10,gregory_2016}. Once R-IMPS reach the ZAMS as fully-radiative AB stars, we expect intrinsic X-ray emission to cease, hence the AB stars in our X-ray bright sample are likely included owing to the presence of relatively bright, unresolved convective companions (themselves IMPS or, more frequently, TTS).

\begin{figure}[ht]
    \centering
    \includegraphics[scale=0.46]{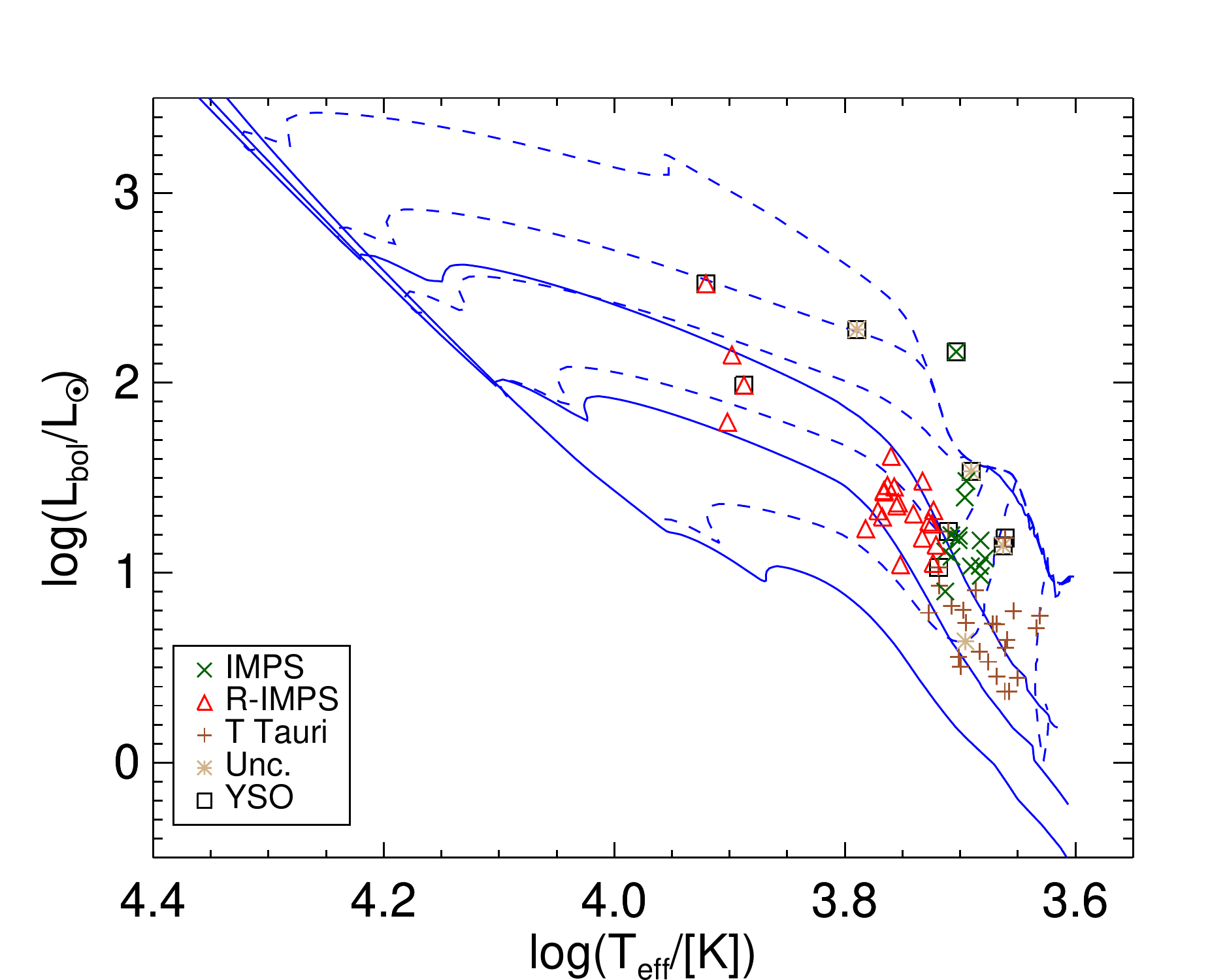}
    \caption{H-R diagram for 65 X-ray bright sources with spectroscopically measured $T_{\rm eff,S}$ from D17 constraining the SED fits. \citet{haemmerle_2019} isochrones (0.1, 1, 3, and 10~Myr) and evolutionary tracks (1, 2, 3, 4, 5 and 7~$M_{\odot}$) are plotted as dashed and solid lines, respectively. YSOs fit using the \citet{robitaille17} models are boxed in black and tend to be found near (and in one instance above) the intermediate-mass stellar birthline, where the various evolutionary tracks overlap. 
    \label{fig_D17HRD}
    }
\end{figure}

In Figure \ref{fig_D17HRD} we plot on an HR diagram the subset of 65 sources from our X-ray bright sample with photospheric temperatures $T_{\rm eff,S}$ measured spectroscopically as part of the Gaia-ESO survey. This subset includes 23 R-IMPS, 14 IMPS, 24 TTS, and 4 sources that remained unclassified due to poorly-constrained SED model parameters. No AB stars are included because D17 did not report $T_{\rm eff,S}$ values for hot stars. For this plot we refit the SEDs of the nine IR-excess sources (boxed) using the updated sets of YSO models provided by \citet{robitaille17}. We then computed $T_{\rm eff}$ and $L_{\rm bol}$ using the mean parameter values for all SED models with $T_{\rm eff}$ falling within 3-sigma of $T_{\rm eff,S}$ reported by D17. In 12 instances (all non-excess sources) no SED models fell within this range, most likely due to the limitations of the \citet{Kurucz} stellar atmosphere models, which do not accurately reproduce the H- opacity minimum at 1.65~$\mu$m for PMS stars with $T_{\rm eff}=4000$--5000~K. In these cases we plotted $T_{\rm eff,S}$ and the mean $L_{\rm bol}$ from the 10 SED models with the nearest $T_{\rm eff}$ parameter values. 

As Figure~\ref{fig_D17HRD} illustrates, R-IMPS (red triangles) are traversing the radiative Henyey tracks and have generally higher L$_{bol}$ compared to fully-convective IMPS (green asterisks). Most R-IMPS and IMPS have stellar masses of 2--4~$M_{\odot}$, but nearly all of the IMPS in this mass range lie above the 1~Myr isochrone, while nearly all the R-IMPS fall between the 1 and 3~Myr isochrones. The lower-mass TTS span the same range in isochronal ages as the R-IMPS and IMPS combined, in good agreement with the ${\le}3$~Myr duration of star formation in the central regions of the Carina Nebula reported by P19. The YSOs (boxed) lie  close the intermediate-mass stellar birthline (defined by the upper-right boundary of the \citealp{haemmerle_2019} evolutionary tracks), including three with masses exceeding $4~M_{\odot}$.



\subsection{Trends in X-ray Luminosity with Stellar Mass and Luminosity}

Up to this point we have focused primarily on comparing the distributions of various X-ray emission properties among our different stellar $TM$ classes, and we have demonstrated that X-ray luminosity is by far the most distinguishing characteristic of (R-)IMPS compared to lower-mass TTS. To investigate trends in X-ray luminosity with the parameters of individual stars, we are limited by the source selection requirements for reliable X-ray spectral fitting. The luminosity range over which our X-ray bright sample appears mostly complete spans ${\sim}1.5$~dex (Figure~\ref{fig_lum_cuml}), which is comparable to the typical scatter observed in stellar $L_X/L_{\rm bol}$ relations \citep{preibisch_2005,gagne_2011}.

To probe more extended trends in our analysis we add another sample of X-ray emitting pre-main-sequence stars to our analysis. The MYStIX (Massive Young Star-Forming Complex Study in Infrared and X-ray) sample is a multi-wavelength survey similar to CCCP that combines \textit{Chandra} X-ray data and NIR/MIR data from 2MASS, UKIRT, and \textit{Spitzer}/IRAC \citep{feigelson_2013}. We only used sources from the CCCP-MYStIX sample that had published X-ray luminosities and for which we could classify their IR counterparts with confidence. We made no cut to the X-ray detection significance or X-ray net counts for these sources.

\begin{figure}[ht]
    \centering
  \includegraphics[scale=0.55]{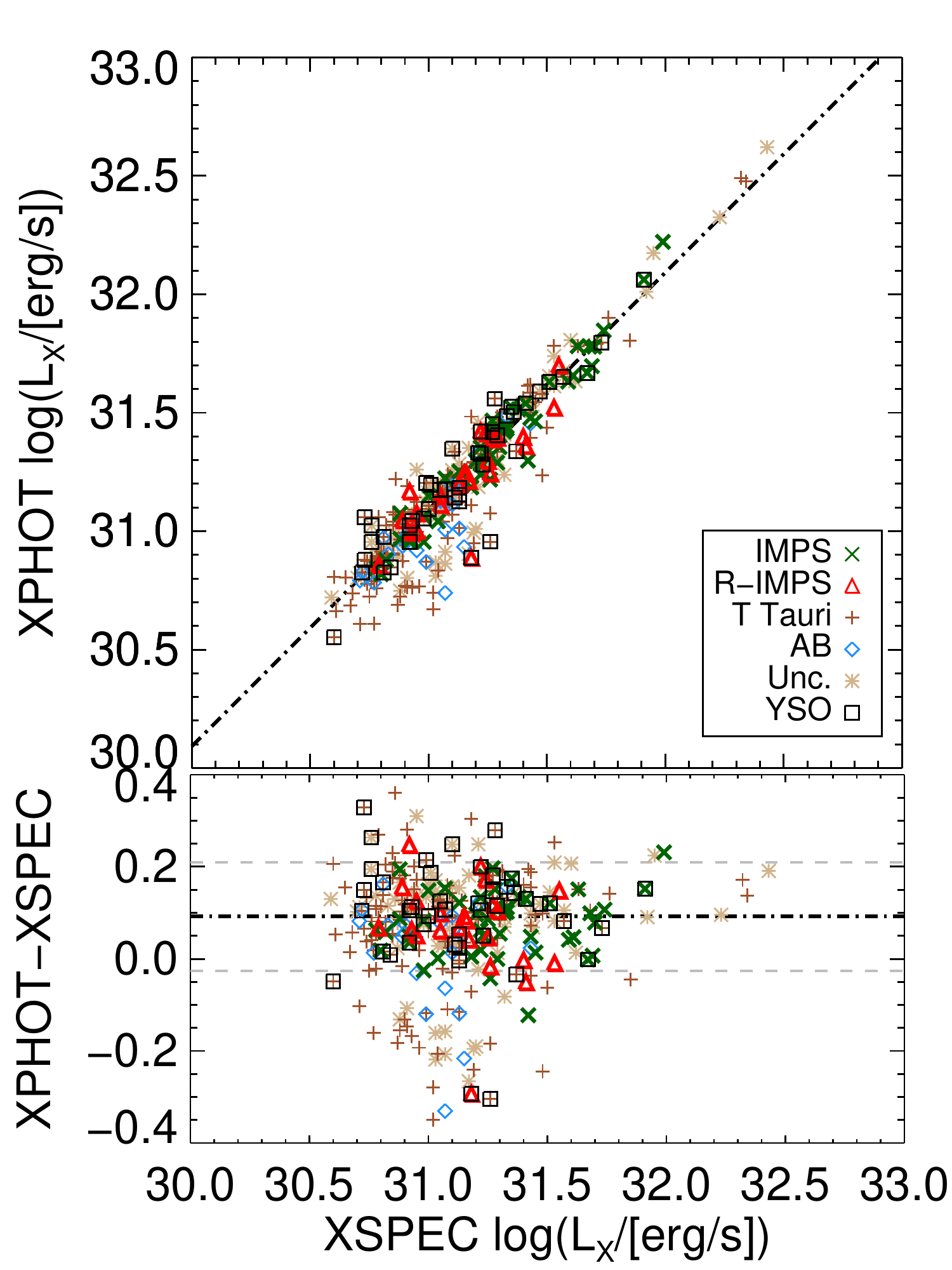}
    \caption{
    \textit{Top:} Total-band (0.5--8~keV) $L_X$ derived from \texttt{XPHOT} \citep{getman+10} plotted against our $L_X$ derived from \texttt{XSPEC}.  The plot legend denotes the $TM$ class of each star; boxes additionally mark YSOs with IR excess emission. The black dash-dotted line shows the 1:1 relation, not a fit to the data. \textit{Bottom:} Ratio (logarithmic difference) between \texttt{XPHOT} and \texttt{XSPEC} $L_X$ values. The black dash-dotted and gray dashed lines denote the mean and ${\pm}1\sigma$ about the mean, respectively.
    \label{fig_xphot_xspec}
    }
\end{figure}

In the MYStIX catalog \citep{MYStIX_MPCM}, reported $L_X$ values were computed using the \texttt{XPHOT} method for faint X-ray sources \citep{getman+10}, which assumes a (luminosity-dependent) template TTS spectral shape to estimate the absorption affecting the observed spectrum.
Because \texttt{XPHOT} was developed empirically using low-mass TTS samples, there is a danger that undiagnosed trends in X-ray spectral shape with stellar mass could bias this method toward systematically different $L_X$ values when applied to IMPS. To check for such systematics, in Figure \ref{fig_xphot_xspec}
we plot  $L_X$ derived from \texttt{XPHOT} (rescaled to the revised Carina distance) versus our spectral fitting (\texttt{XSPEC}) for the sources in common among our sample and MYStIX.
We find generally good agreement, with $L_X$ within 0.1 dex, between the two methods across our sample,
with the large majority of sources falling within ${\sim}0.1$ dex of the 1:1 line between the two methods. There are no evident trends among the various $TM$ classes. We therefore can reliably compare the $L_X$ from either method and incorporate the larger, MYStIX sample to probe trends in $L_X$. 

 G16 adopted previously-published X-ray luminosities derived from \texttt{XPHOT}, and the generally good agreement we have found between \texttt{XPHOT} and our more careful \texttt{XSPEC} analysis (Figure~\ref{fig_xphot_xspec}) reinforces their conclusions about the decay of X-ray luminosity with time.

\begin{figure}
    \centering
    \includegraphics[scale=0.45]{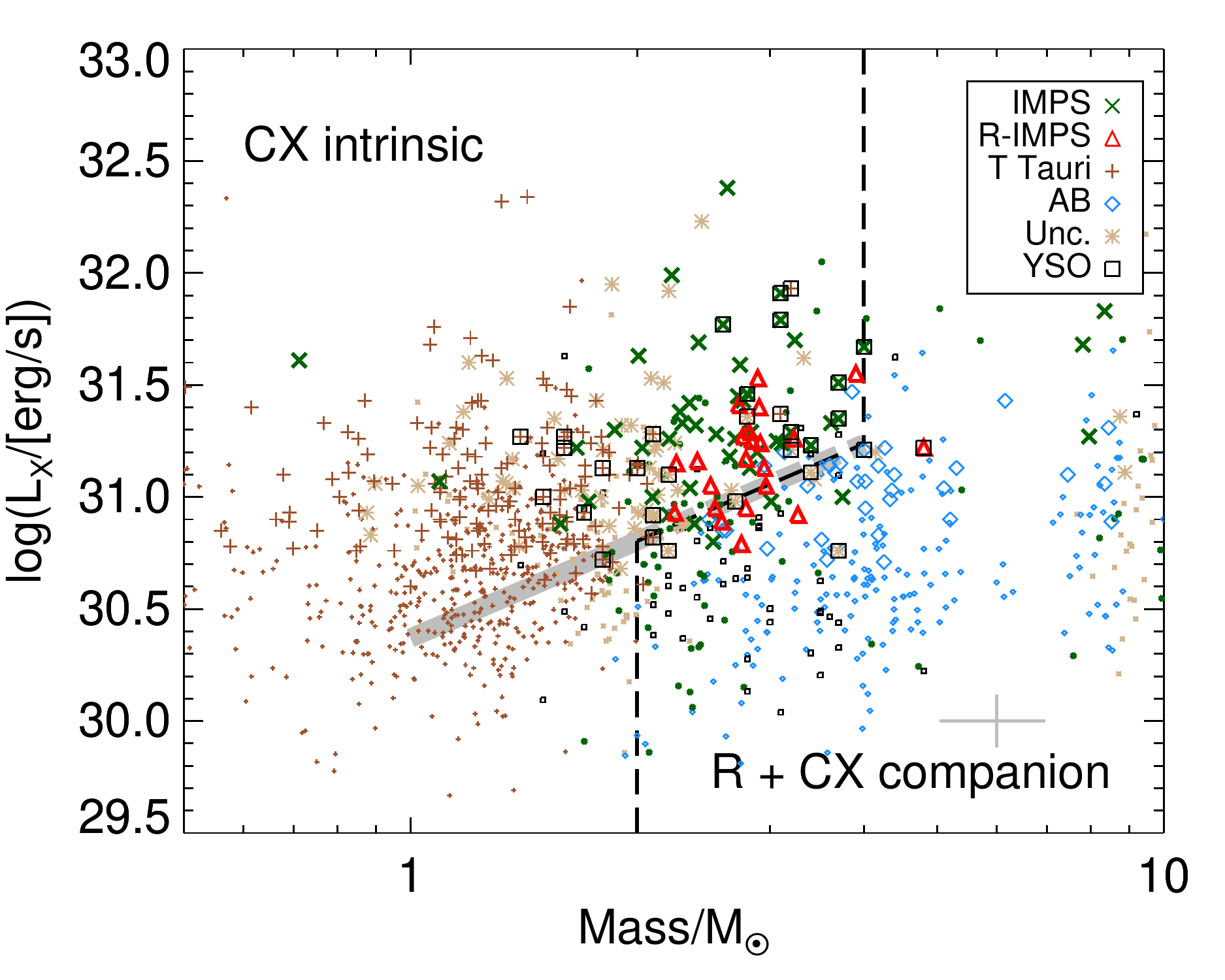}
    \caption{Total-band, absorption-corrected $L_X$ versus stellar mass from IR SED fitting for 310 sources in our \texttt{XSPEC} bright sample (large symbols) with well-constrained masses from SED fitting plus 1264 fainter X-ray sources with \texttt{XPHOT} luminosities reported in the CCCP--MYStIX catalogs (smaller symbols). Boxes denote YSOs with masses reported by \citet{povich_2011}. The solid gray bar shows the $L_X$-Mass linear regression for low-mass T Tauri stars in the COUP sample (P05), and the dashed gray bar extrapolates this trend from 2--4 $M_{\odot}$. The dashed black line segments roughly separate stars with at least partially-convective envelopes and powering intrinsic, coronal X-ray emission (CX intrinsic) from stars with radiative envelopes and X-ray emission likely produced by lower-mass, convective companions (R+CX companion). 
    The grey cross at the bottom right of the plot shows the median error bar for the plotted points.
    \label{fig_xlum_mass}
    }
\end{figure}

In Figure \ref{fig_xlum_mass} we plot $L_X$ against stellar mass derived from IR SED fitting. We broadly split the figure into two regions: sources powered by intrinsic coronal X-ray (CX) emission consistent with a convective interior versus sources whose IR bright primary star has a mostly or fully radiative envelope (R), meaning the X-ray emission most likely originates from a lower-mass, convective TTS (or IMPS) companion (CX companion), that is unseen in the IR. Some magnetic ``peculiar" (Ap) stars have been detected in X-rays \citep[][]{stelzer_2003} but these would be far too faint to be detected at the Carina distance, given CCCP sensitivity limits . 
These dividing lines are based on the P05 $L_X$ vs M$_\star$ linear regression fit extrapolated for 2 M$_{\odot}$ to 4 M$_{\odot}$ stars. We find that the majority of IMPS and even R-IMPS fall above this extrapolation line, while the vast majority of AB X-ray sources fall below it.

\begin{figure}
    \centering
    \includegraphics[scale=0.45]{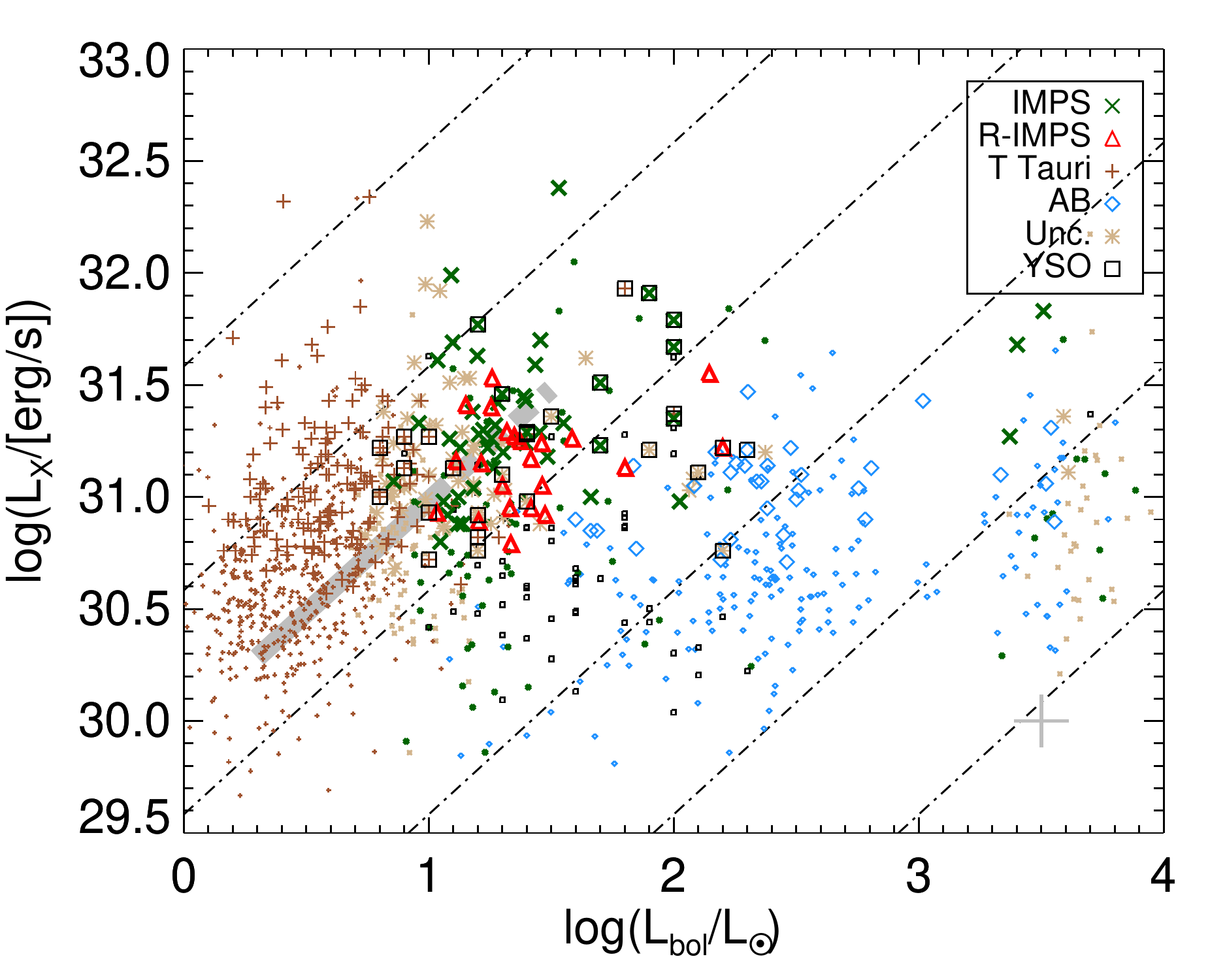}
    \caption{$L_X$ versus $L_{bol}$ for the same sources plotted in Figure~\ref{fig_xlum_mass}.  The solid gray bar shows the P05 $L_X$--$L_{bol}$ relation for the COUP sample, and the dashed grey bar extrapolates this trend from 2--4~$M_{\odot}$. Dash-dotted black lines show loci of constant $\log{(L_X/L_{\rm bol})} = -2,~-3,~-4,~-5,~-6$, and $-$7. YSOs from \citet{povich_2011} had $L_{\rm bol}$ reported to a precision of 0.1~dex. The gray cross at bottom-right shows the median error bars on the plotted points.
    \label{fig_lumbol_xlum}
    }
\end{figure}

This separation in $L_X$ between IMPS, R-IMPS, and AB stars of 2--4~$M_{\sun}$ illustrates the evolution of X-ray emission as a probe of IMPS interior structure: IMPS begin as (fully) convective and X-ray luminous, then develop a radiative zone and become R-IMPS, at which point the X-ray luminosity declines. Finally, a fully-radiative envelope completes the transition to an AB star, spelling the end of magneto-coronal X-ray emission. At this point an X-ray quiet AB star would drop out of our sample, unless it has a lower-mass X-ray bright companion. This means that the AB sources have X-ray emission characteristic of TTS, not IMPS, dropping them below the extrapolated P05 $L_X$--mass relation. The MYStIX sample allows us to include many more of these AB sources with lower $L_X$ in Figure~\ref{fig_xlum_mass} (smaller symbols) than appear in our X-ray bright sample.

In Figure \ref{fig_lumbol_xlum} we plot $L_X$ against $L_{bol}$ derived from IR SED fitting. 
Both IMPS and R-IMPS generally follow the TTS log(L$_X$/L$_{bol}$) relation, with the large majority falling in the range of  $-4<\log(L_X/L_{\rm bol})<-3$. The AB sources do not follow this relation because $L_X$ and $L_{\rm bol}$ are decoupled in the case of unresolved binary systems where the IR and X-ray emission are dominated by different stellar components, as discussed above. No sources in Figure~\ref{fig_lumbol_xlum} fall below the $\log(L_X/L_{\rm bol})= -7$ line, indicating that few to none of the AB stars exhibit intrinsic X-ray emission from shocked stellar winds \citep[Carina OB stars have $\log(L_X/L_{\rm bol})= -7.26\pm 0.21$;][]{naze+11}.


\begin{figure}[htp]
    \centering
    \includegraphics[scale=0.2]{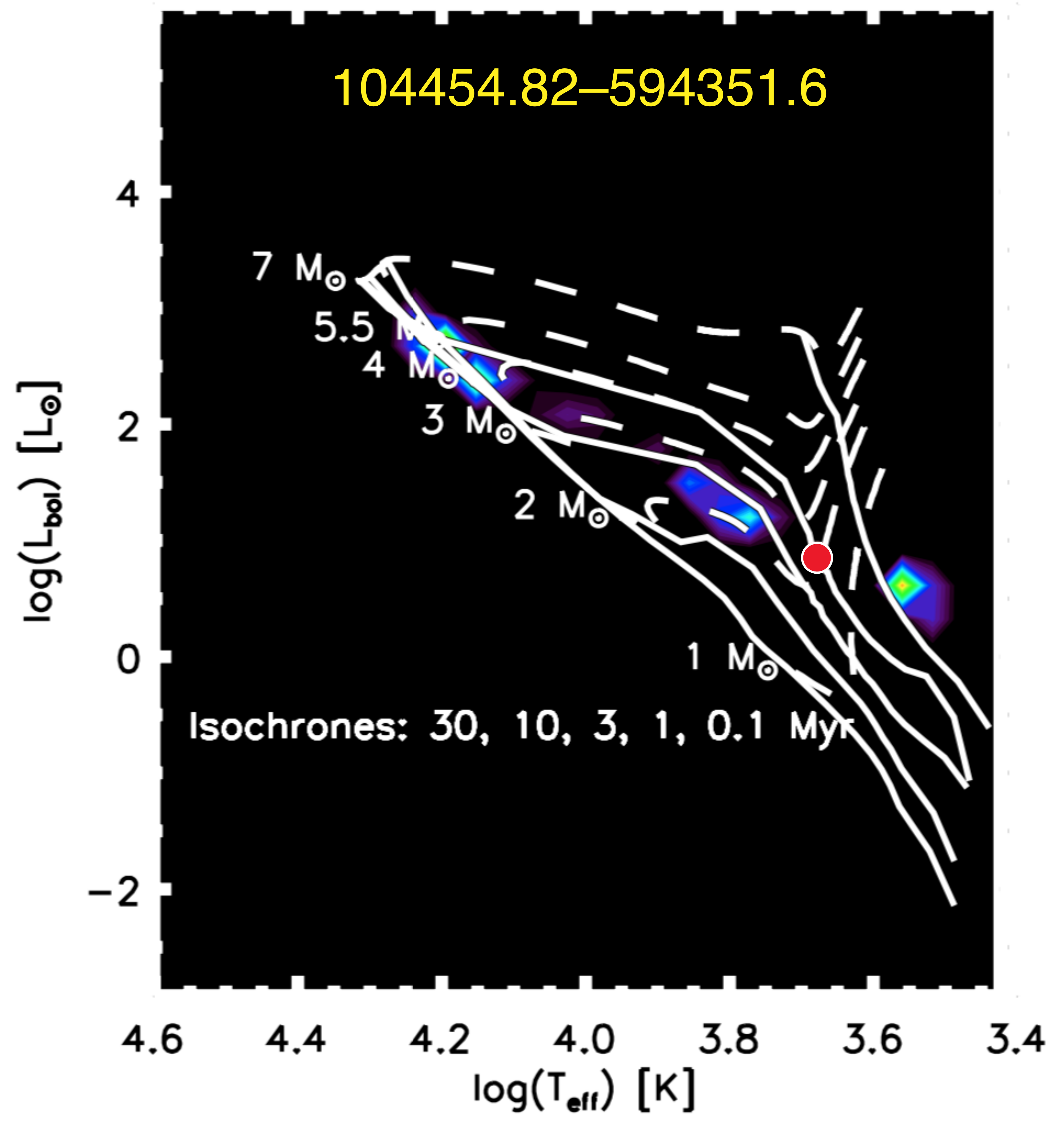}
    \includegraphics[width=0.9\columnwidth]{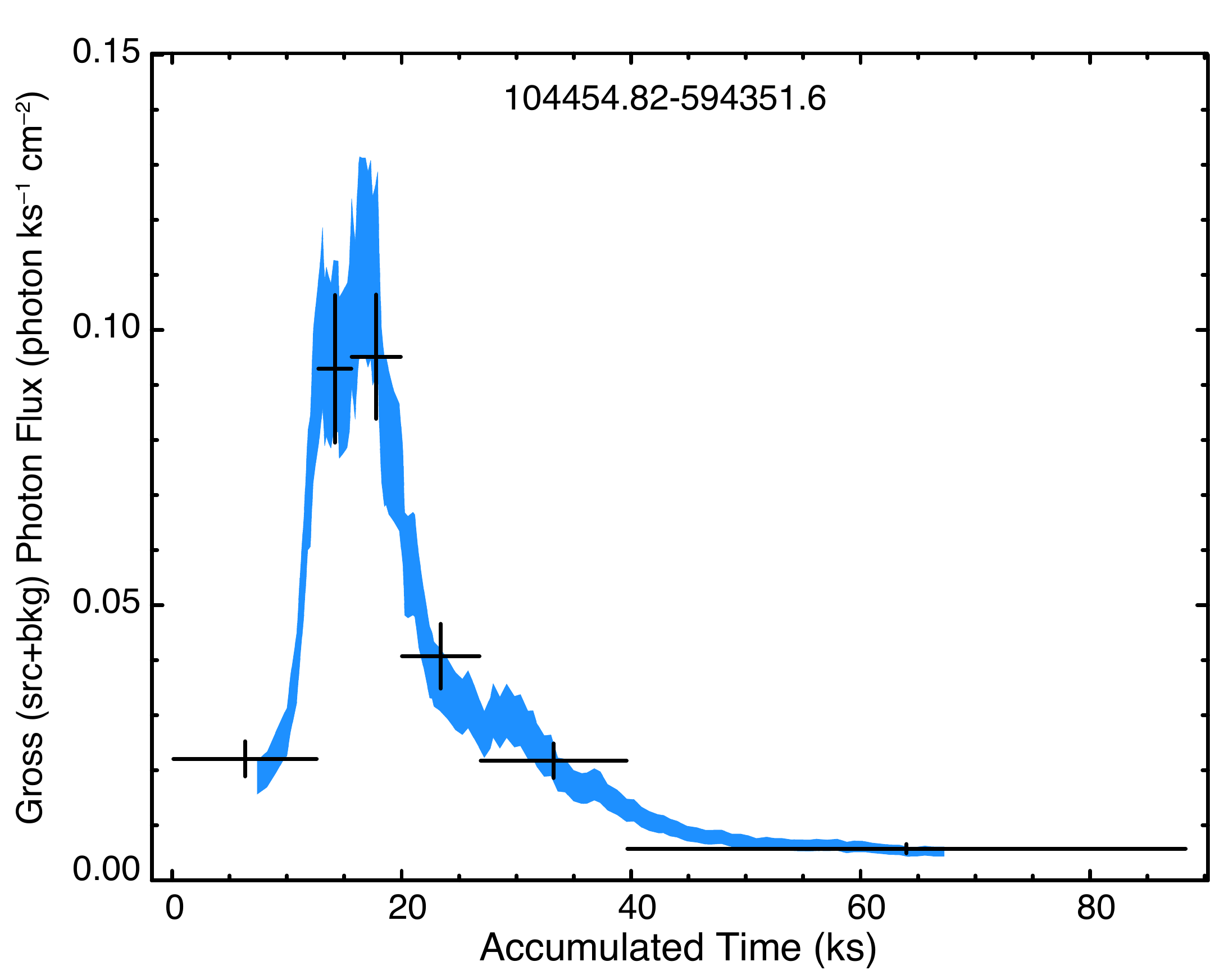}
    \includegraphics[width=0.9\columnwidth]{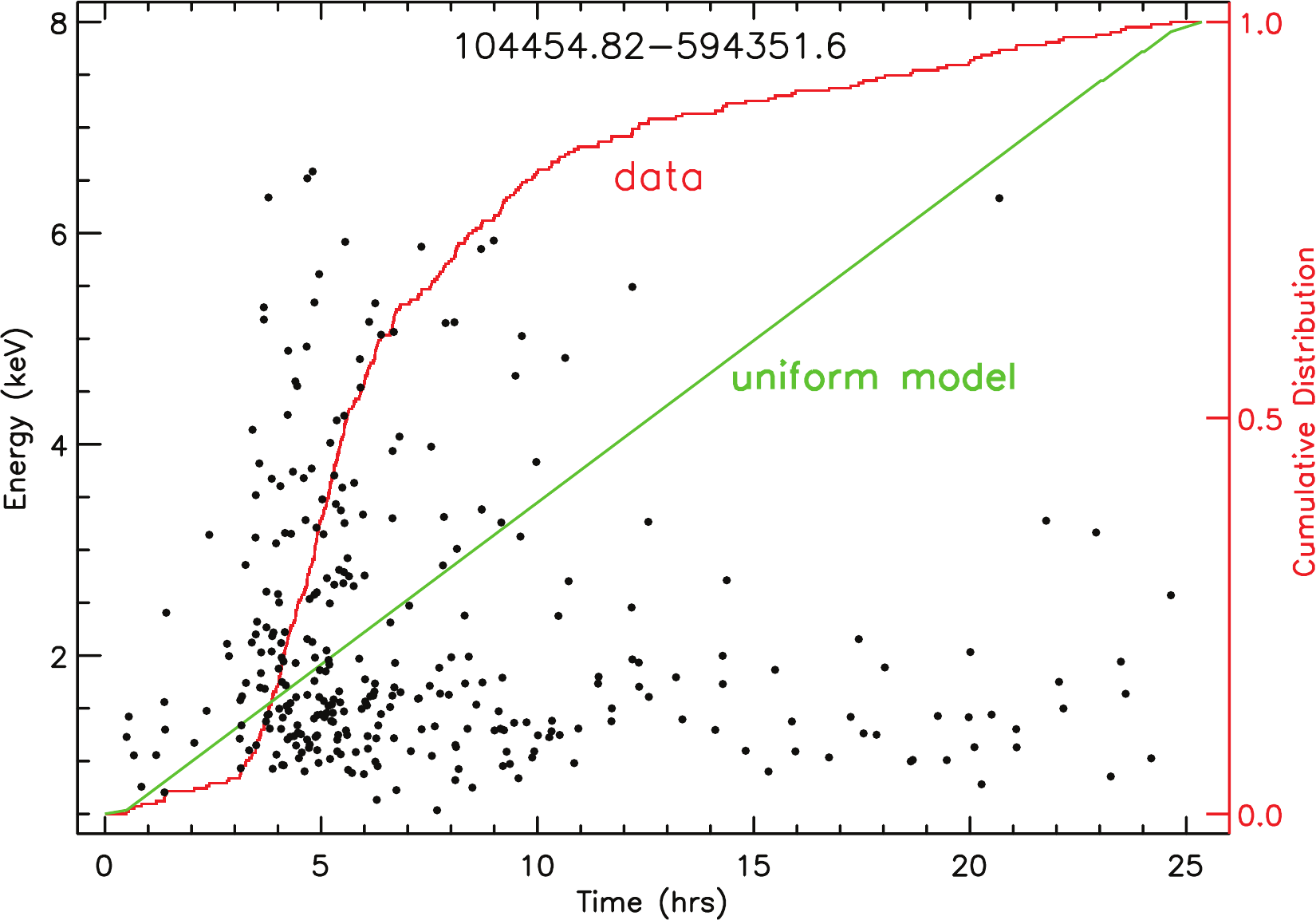}
    \caption{\textit{Top:} Individual pHRD for the IR counterpart to CXOGNC J104454.82-594351.6 (Section \ref{sec_5IMPS_unc_mostxluminous}). \citet{siess_2000} isochrones and evolutionary tracks are displayed as in Fig.~\ref{fig_xspec_phrd}, with a red dot marking the precise location of the star using $T_{\rm eff,S}$ measured from spectroscopy.
    \textit{Middle:} Light curve for the X-ray counterpart to CXOGNC J104454.82-594351.6 displayed as smoothed uncertainty envelope (blue) of photon arrival rate and grouped data with error bars (black) demonstrating X-ray variability.
    \textit{Bottom:} Photon arrival diagram for the X-ray source. Points show arrival time and median energy of individual events, while the red and green curves show the cumulative distribution of arrival times for data and uniform models, respectively. 
    \label{fig_5IMPS_unc_mostxluminous}
    }
\end{figure}

\vspace{1em}
\subsection{Interesting Individual IMPS}\label{sec_interesting_imps}

In this section we discuss nine individual stars and YSOs drawn from the 77 IMPS and R-IMPS candidates in our sample. 
All of these featured objects have  $T_{\rm eff,S}$ measured by D17, increasing the precision of their locations on the HR diagram (Figure~\ref{fig_D17HRD}). Table \ref{tab_best_parameters} summarizes the best-fit X-ray parameters for all sources in our sample, with the 9 sources discussed in this section highlighted.

\subsubsection{The Brightest X-ray Flare from a Spectroscopically Classified, Diskless IMPS} \label{sec_5IMPS_unc_mostxluminous}
Among the IMPS for which we have spectroscopic classifications, CXOGNC J104454.82-594351.6 is the most X-ray luminous, with $\log(L_{X}) = 31.74$~erg~s$^{-1}$. Nearly 60\% of this luminosity is emitted in the hard band (${>}2$~keV).
Without spectroscopy, this IMPS would remain unclassified by our $TM$ scheme, because 
the naked PMS model fits to its SED were unconstrained (top panel of Figure~\ref{fig_5IMPS_unc_mostxluminous}). The H$^-$ opacity bump likely caused the SED fits to miss the correct $T_{\rm eff,S} = 4768$~K (D17), which gives a mass of ${\sim}2~M_{\odot}$. This star therefore barely exceeds our mass threshold for IMPS classification, and its high X-ray luminosity appears to be due to variability produced by one or more large flares. 
The middle panel of Figure \ref{fig_5IMPS_unc_mostxluminous} shows the X-ray light curve of CXOGNC J104454.82-594351.6, which is dominated by a fast-rise and exponential-decay flare \citep{COUP_Flares,getman+2021}. \edit1{The blue curve is the adaptively smoothed uncertainty envelope of the observations \citep{broos_2010} which is smoothed to produce bins with SNR = 6.0 and 
whose thickness represents the error bars in the photon flux.}


\subsubsection{Two IMPS Near the Maximum Mass for Fully-Convective, Non-Accreting Stars}\label{sec_5IMPS_YSO_mostxluminous}
\begin{figure}
    \centering
        \includegraphics[width=\columnwidth]{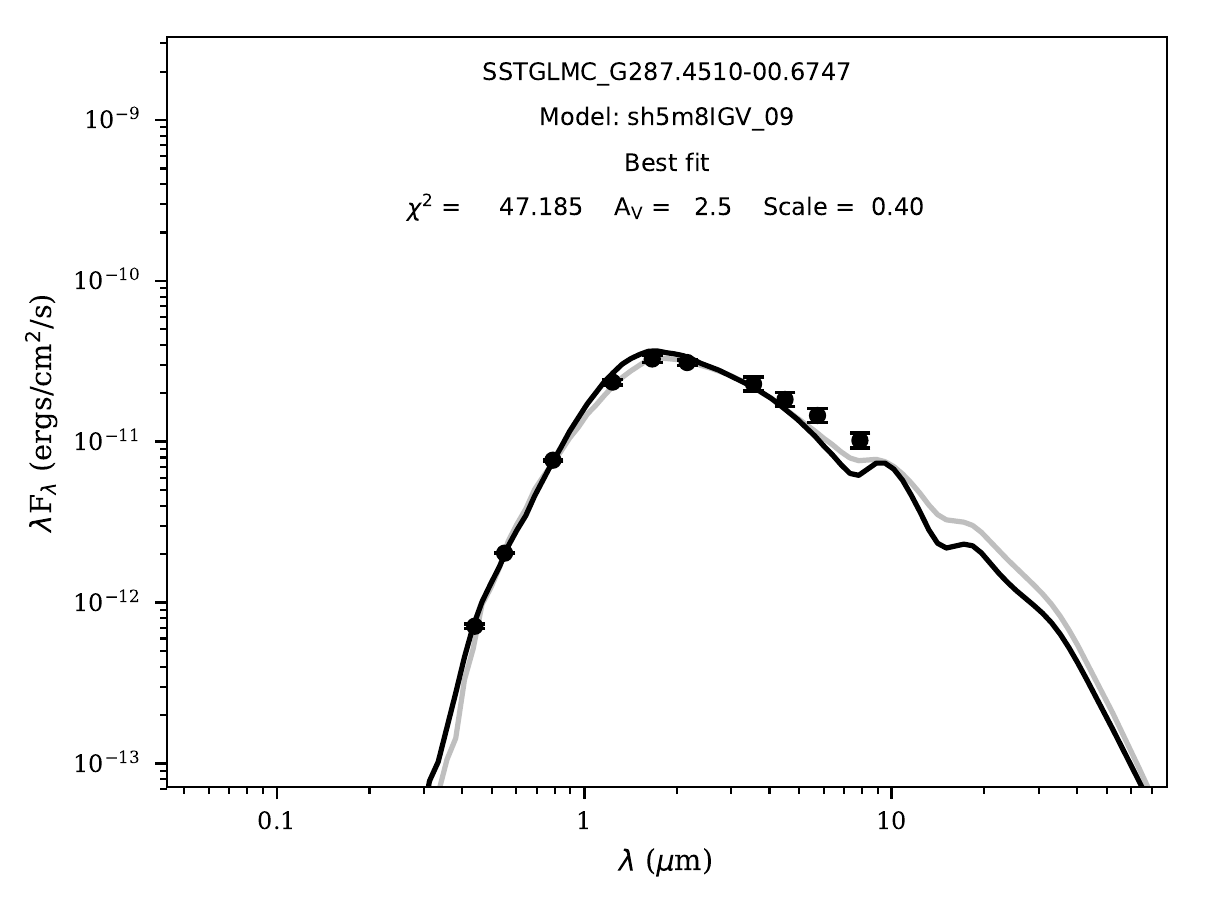} \\
    \includegraphics[width=\columnwidth]{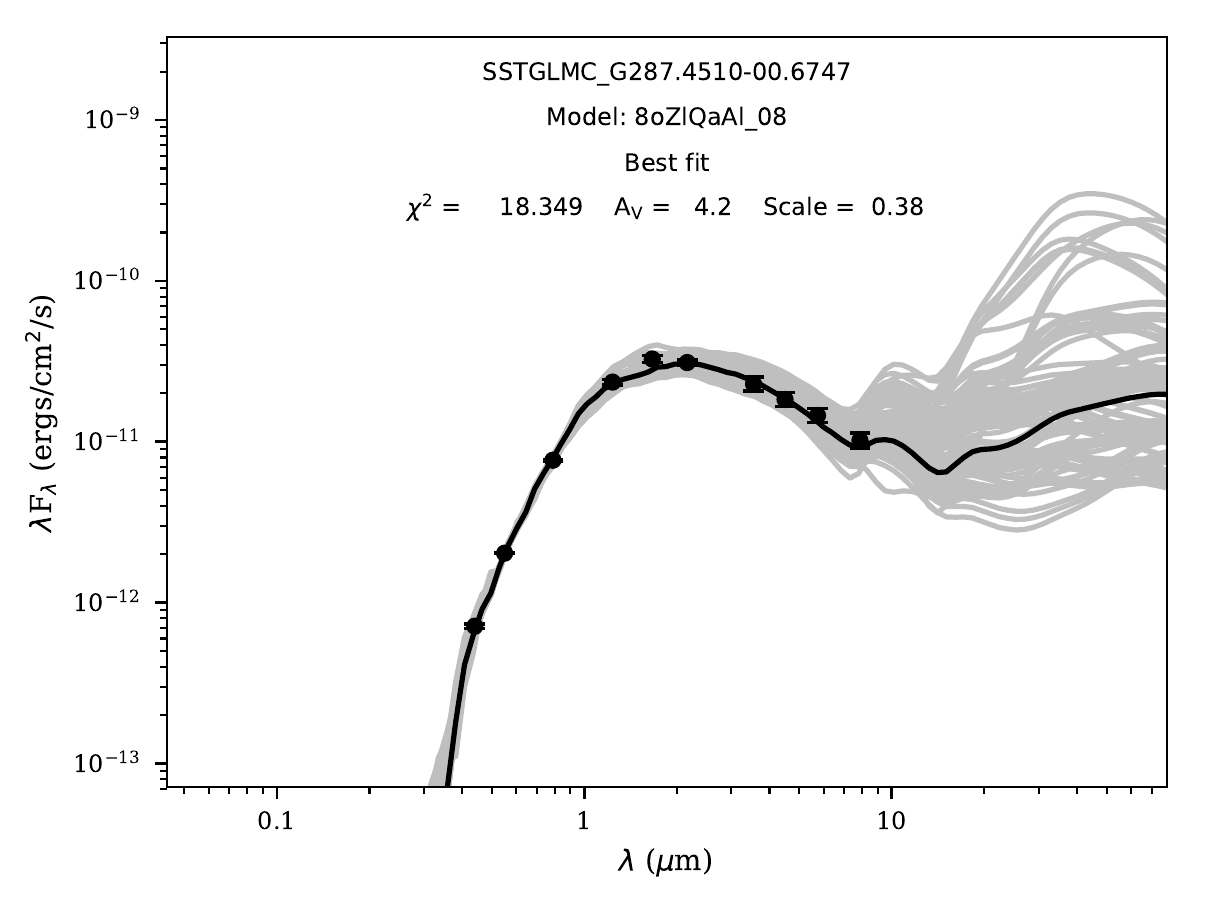}
    \caption{YSO model fits \citep{robitaille17} to the visible through mid-IR SED associated with CXOGNC  J104352.31-593922.2. \textit{Top:} The best two disk-only models do not reproduce the strength and shape of the observed IR excess emission. \textit{Bottom:} Models including both a disk and an infalling envelope with a bipolar cavity provide a large number of acceptable fits. Thermal IR photometry at ${>}10$~\um\ wavelengths would be required to further constrain the fits (but this would be challenging given the bright IR nebular background in the Carina Nebula).
    }
    \label{fig_5IMPS_YSO_mostxluminous}
\end{figure}

CXOGNC J104352.31-593922.2 appears to be the least-evolved YSO among the 65 stars in our spectroscopic subsample. It has a high $\log(L_X) = 31.35$~erg~s$^{-1}$ with no evident variability. 
 We incorporated available $BVI$ photometry from D17 into the IR SED and refit with both disk-only and disk+infalling envelope YSO models from \citet{robitaille17}. Example plots illustrating the multiple acceptable model fits to the photometry data are shown in Figure \ref{fig_5IMPS_YSO_mostxluminous}, and it is clear that the disk+envelope models provide qualitatively better fits to the MIR SED.

The best SED models for this YSO that agree with the correct $T_{\rm eff,S} = 5126$~K (D17) indicate a stellar mass of 3--4~$M_{\odot}$ and a position very near the \citet{haemmerle_2019} birthline. 
This interpretation is consistent with the presence of a circumstellar disk and envelope, indicating the possibility of recent or ongoing mass accretion. Even in a flare, such high X-ray luminosity is unlikely to be produced by an unresolved, lower-mass T Tauri companion, making CXOGNC J104352.31-593922.2 one of the youngest and most massive YSOs powering coronal X-ray emission that we would expect to discover, 
given its location at or above the maximum mass for the fully-convective birthline \citep{haemmerle_2019}.\footnote{CXOGNC J104352.31-593922.2 appears to lie above the birthline in Figure~\ref{fig_D17HRD}, but that is an artifact of the averaging of parameters over multiple models, since the majority of the well-fit YSO models for this source prefer temperatures hotter than $T_{\rm eff,S}$, with commensurately higher bolometric luminosities.} 

\begin{figure}
    \centering
    \includegraphics[scale=1.6]{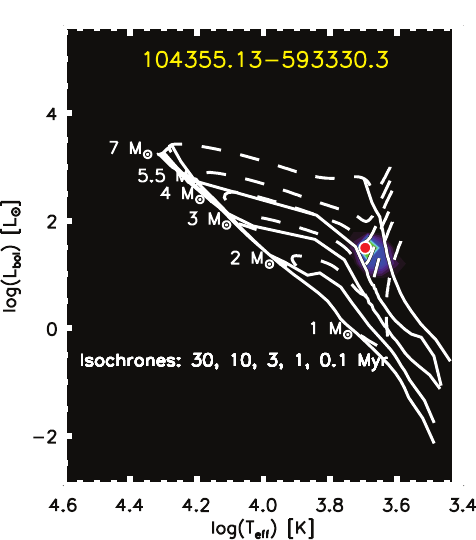}
    \includegraphics[scale=0.3]{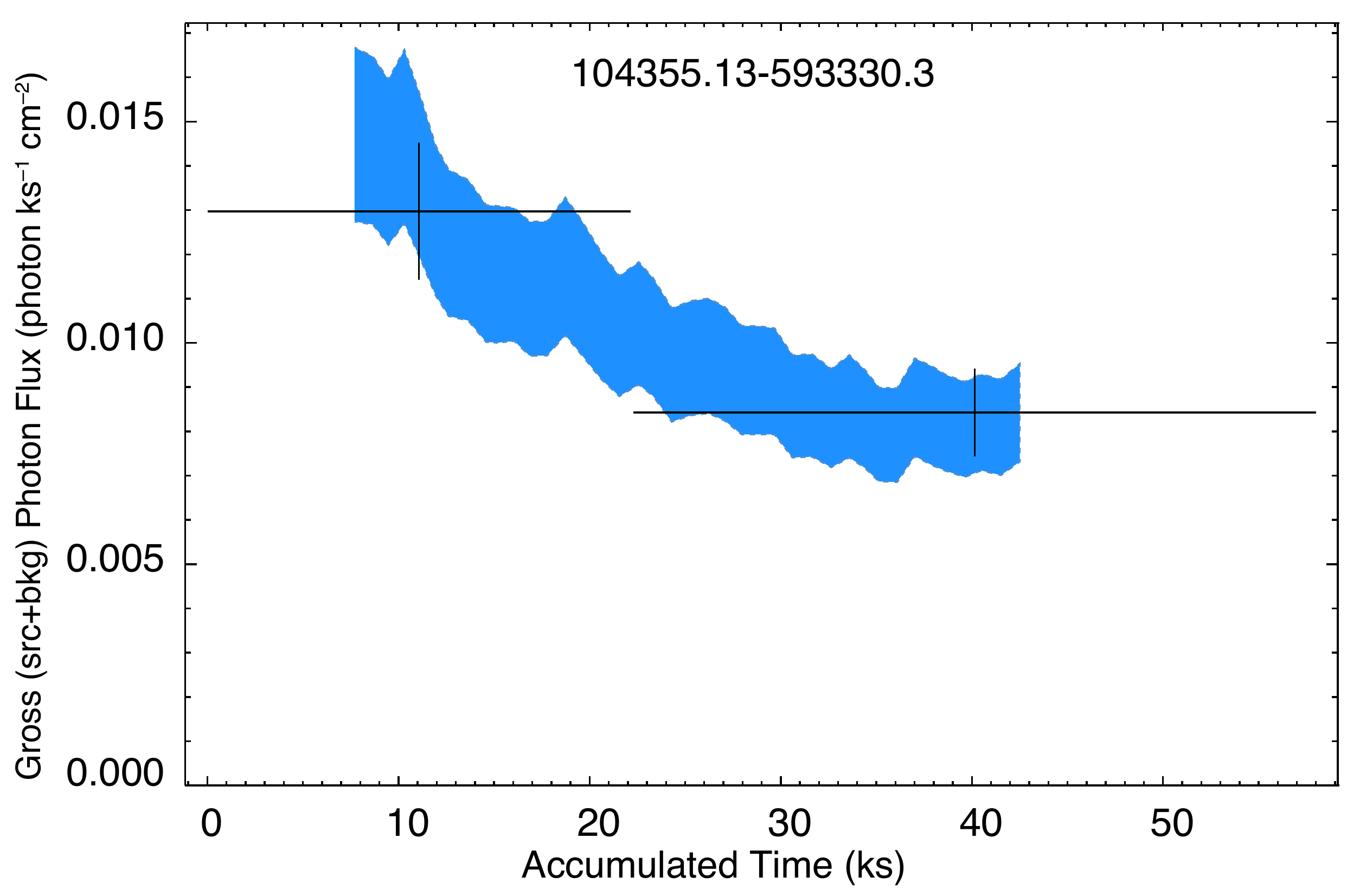}
    \caption{\textit{Top:}Individual pHRD for the IMPS associated with CXOGNC J104355.13-593330.3. Overlays are the same as in Figure~\ref{fig_5IMPS_unc_mostxluminous}.
    \textit{Bottom:} X-ray light curve, with colors and symbols the same as in Figure~\ref{fig_5IMPS_unc_mostxluminous}. 
    \label{fig_5IMPS_mostmassivediskless}
    }
\end{figure}

 A second IMPS (CXOGNC J104355.13-593330.3) also has a high X-ray luminosity ($\log{L_X}=31.41$~erg~s$^{-1}$) and a remarkably high mass. It is possibly variable (Section \ref{sec_analysis}), with a lightcurve showing what could be post-flare exponential decay over the 60-ks observation (Figure~\ref{fig_5IMPS_mostmassivediskless}). It presents a cooler X-ray spectrum than most stars in our sample, with 67\% of emission at softer energies (${<2}$~keV). Its MIR counterpart shows no excess emission out to 4.5~\um.  SED modeling with stellar atmospheres indicates a mass ${>}3~M_{\odot}$, making CXOGNC J104355.13-593330.3 an example of the most massive, fully-convective, non-accreting IMPS theoretically allowed by \citet{haemmerle_2019}. The SED fit parameters found excellent agreement with $T_{\rm eff,S}=4994$~K (D17), and the X-ray spectrum was best fit with the 1T Free models. 

This extreme star is located close to the heart of Tr~14, the densest cluster in Carina (Figure \ref{fig_carina}) with many surrounding stars within 3\arcsec. It is possible that its high inferred luminosity, and hence mass, could result from the blending of two (or more) sources. This could more strongly affect the IR SED fits than the X-ray spectral fits, since \textit{Chandra} has a higher (on-axis) spatial resolution compared to \textit{Spitzer}/IRAC or 2MASS. 


\subsubsection{Two Diskless IMPS at the Convective-Radiative Transition
} \label{sec_5IMPS_radiativeimps}

\begin{figure}
    \centering
    \includegraphics[scale=1.5]{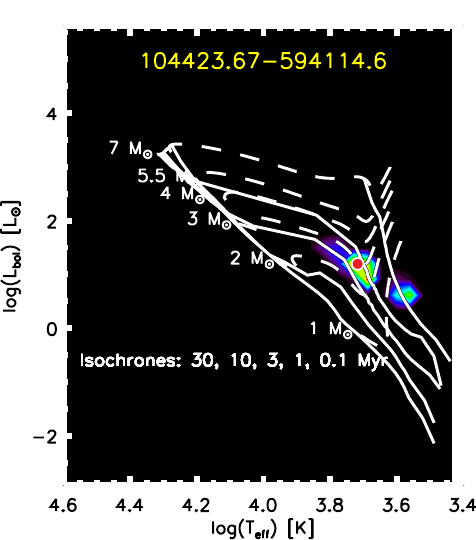}
    \includegraphics[scale=1.5]{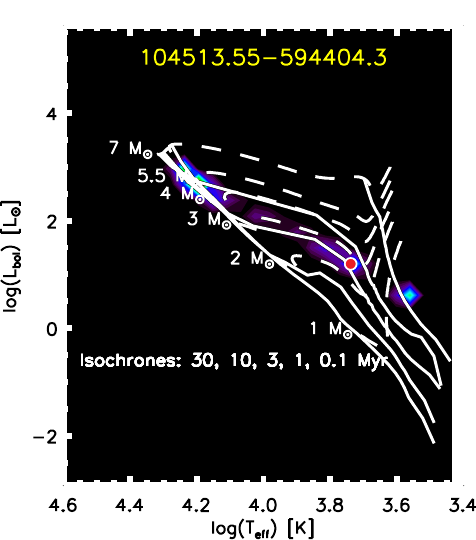}
    \caption{Individual pHRDs for the IR counterparts to CXOGNC J104423.67-594114.6 (top) and J104513.55-594404.3 (bottom). 
    Overlays are the same as in Figure~\ref{fig_5IMPS_unc_mostxluminous}.
    \label{fig_5IMPS_radiativeimps}
    }
\end{figure}

CXOGNC J104423.67-594114.6 and J104513.55-594404.3 are both diskless IMPS with $\log( L_X)\la 31$~erg~s$^{-1}$, on the faint end of the (R-)IMPS in our sample. 
While their SED fit parameters were too poorly constrained for $TM$ classification without spectroscopy, they fell on either side of our (admittedly blurry) temperature dividing line between IMPS and R-IMPS, with $T_{\rm eff,S}= 5218$~K  and 5515 K (D17), respectively (Figure~\ref{fig_5IMPS_radiativeimps}). It is possible that the X-ray emission from each of these two IMPS has begun to decay due to the development of an interior radiative zone. 

\subsubsection{The Four Hottest R-IMPS} 
Four of the R-IMPS plotted in Figure~\ref{fig_D17HRD} have $T_{\rm eff,S}>7500$~K (D17), making them late A-type stars with fully radiative envelopes.\footnote{An argument could be made to re-classify these stars as AB, but we have reserved this classification for hot, intermediate-mass stars on or near the ZAMS.}
Two of these, CXOGNC J104401.10-593535.1 and J104538.35-594207.5, exhibit both IR excess and H$\alpha$ emission, making them candidate Herbig Ae stars. These two YSOs and the diskless, hot R-IMPS CXOGNC J104432.58-593303.5 have $\log{L_X}<31.2$~erg~s$^{-1}$, which is below the median X-ray luminosity of both the TTS and AB classes. Their observed X-ray emission is consistent with convective, lower-mass TTS companions. 

The fourth star in this group, CXOGNC J104446.53-593413.3, has $\log{L_X}=31.55$~erg~s$^{-1}$. This is the most massive of the diskless R-IMPS in our sample, at ${\sim}4~M_{\odot}$ (Figure~\ref{fig_D17HRD}). Assuming this star is fully-radiative and hence incapable of producing magneto-coronal X-rays, the high $L_X$ is most likely powered by a companion that is itself a fully-convective IMPS of 2--3~$M_{\odot}$. 

%

\begin{table*}[htb]
\centering
\scriptsize
\caption{\edit1{Best Fit X-ray Parameters of Noteworthy IMPS}} \label{tab_best_parameters}
    \begin{tabular}{cccccccccc}
    \hline
    \hline
    XName   &TMClass   &YSO   &FrozenModel   &NetCounts   &Variable  &NHX   &kT1   &logLx\_tc   &logLbol   \\
   &   &   &   &(Counts)   &   &($10^{22}\;\rm{cm}^{-2}$)    &(keV)   &($\rm{erg} \; \rm{s}^{-1}$)   &($\rm{L}_\odot$)   \\
\hline
104454.82-594351.6&IMPS&n&n&300&2&$0.41_{-0.24}^{+0.61}$&$5.4^*$&31.7&1.7   \\
104352.31-593922.2&IMPS&y&n&139&0&$0.81_{-0.42}^{+1.42}$&$2.7_{-1.7}^{+5.0}$&31.4&2.0   \\
104355.13-593330.3&IMPS&n&n&140&2&$0.65^*$&$1.5_{-1.3}^{+2.1}$&31.4&1.3   \\
104423.67-594114.6&IMPS&n&n&144&0&$0.14^*$&$1.9_{-1.4}^{+2.7}$&30.9&1.0   \\
104513.55-594404.3&RIMPS&n&n&74&0&$0.17^*$&$3.6^*$&30.8&1.2   \\
104401.10-593535.1&RIMPS&y&n&60&0&$1.21_{-0.51}^{+1.91}$&$0.9_{-0.5}^{+1.4}$&31.1&2.3   \\
104538.35-594207.5&RIMPS&y&n&105&1&$1.14_{-0.53}^{+2.03}$&$4.9^*$&31.2&2.2   \\
104432.58-593303.5&RIMPS&n&n&63&0&$0.44_{-0.11}^{+0.91}$&$2.0_{-1.4}^{+4.2}$&31.1&1.8   \\
104446.53-593413.3&RIMPS&n&n&135&0&$0.29^*$&$2.7_{-2.0}^{+5.4}$&31.5&2.1   \\

    \hline
    \end{tabular}
        \tablenotetext{*}{Upper limit on $N_{\rm H}$ or lower limit on $kT$ parameter.}
\end{table*}

\section{Conclusions} \label{sec_conclusions}

We have analyzed the X-ray emission properties of 370 X-ray bright stars in the Carina Nebula that we classified via IR SED fitting as IMPS, R-IMPS, TTS, or AB stars (Section \ref{sec_src_class}). 
X-ray spectral fitting with thermal plasma emission models 
returned physical parameters.
We found that 82\% of sources had good agreement between the hydrogen absorbing columns determined independently 
using our X-ray spectral fitting and IR SED fitting (or NIR colors in a few instances). 

Convective IMPS are systematically more X-ray luminous, by $\sim$0.3 dex (Section \ref{sec_discussion}), than all other low- and intermediate-mass stars in our sample. The mean total-band (0.5--8 keV) luminosity of ($\log{L_{t,c}}=31.4$)~erg~s$^{-1}$ for IMPS falls within the typical ranges for wind-driven emission from OB stars ($\log{L_{t,c}} = 30.6$ to 31.8; only ${\sim}10\%$ have $\log{L_{t,c}} > 31.8$, \citealp{gagne_2011}). Based on X-ray brightness alone, an IMPS could be mistaken for a late O or early B-type star.  


\deleted{IMPS (and R-IMPS) follow the same log(L$_X$/L$_{bol}$) relation as TTS (Figure \ref{fig_lumbol_xlum}) and have similar $kT$ distributions. 
This combined with their systematically high $\log{L_{t,c}}$ 
suggest that their X-ray emission mechanism must be a scaled up TTS dynamo.}

\deleted{Similarly, IMPS (and R-IMPS) follow both the L$_X$-M$_\star$ and L$_X$-L$_{bol}$ linear regression from P05 when extrapolated to 4 M$_{\odot}$. The wellness of this extrapolation further supports the origin of IMPS X-ray emission being analogous to TTS given that P05 sample was dominated by TTS.}

Spectroscopically-classified R-IMPS on radiative Henyey tracks have lower X-ray luminosity compared to IMPS, but higher than low-mass TTS. Combined with their similar $kT$ distributions to IMPS and TTS, 
this signals that R-IMPS are producing their final breaths of X-ray emission before joining the ZAMS as AB stars. 

AB and TTS in our X-ray bright sample exhibit similar XLFs and plasma temperature distributions (Section \ref{sec_lum}). 
No AB stars exhibit sufficiently low $L_X/L_{\rm bol}$ to be consistent with a wind-driven X-ray emission mechanism.
We conclude that the apparent X-ray emission from AB stars in our sample originates from IR-dim, but X-ray bright, low-mass TTS companions (Section \ref{sec_discussion}).


The preponderance of evidence supports a common magneto-coronal flaring mechanism for coronal X-ray production for IMPS, R-IMPS, and TTS across the wide mass range from ${\sim}4~M_{\odot}$ down to the hydrogen-burning limit (P05). We observe the time-decay of X-ray luminosity with growth of a radiative zone reported by G16, but with a much larger sample of IMPS and more careful determination of absorption-corrected X-ray luminosity. We find that the absorption-corrected luminosity determined by assuming a TTS spectral template (\texttt{XPHOT} method) generally gives accurate results for IMPS and R-IMPS, indirect evidence for similar underlying spectral shapes for all coronal X-ray emission from PMS stars of all masses. \edit1{In a couple of instances, we find evidence for fast-rise, exponential-decay lightcurves in IMPS that resemble well-studied lightcurves of TTS flares \citep{COUP_Flares, getman+2021}.}

The most massive, non-accreting, fully-convective IMPS, as evidenced by both their strong coronal X-ray emission and location on the HR diagram (Figure~\ref{fig_D17HRD}) have masses of 3-- 4~$M_{\odot}$. 
We also find 3 X-ray bright YSOs with spectroscopically-constrained SED models returning higher masses than any diskless IMPS, but IR excess emission indicative of recent or ongoing accretion. These empirical constraints on the location of the intermediate-mass birthline show very good agreement with the evolutionary models of \citet{haemmerle_2019}.

The rapid evolution of stellar interior structure from fully-convective to fully-radiative, as traced by the transition from IMPS to R-IMPS to AB means that IMPS provide sensitive probes of isochronal ages for the first ~10 Myr of a massive star stellar population \citep{povich_2019}. The high concentration of IMPS in the Tr 14 cluster compared to the higher fraction of R-IMPS in Tr 16 (Figure \ref{fig_carina}) indicates that Tr 14 is even younger than Tr 16 (Appendix A14 in \citealp{AgeJX} gives a similar result).

\section*{Acknowledgements}
We thank K. V. Getman and L. A. Hillenbrand for helpful discussions that improved this paper. \edit1{We thank our referee for thoughtful and helpful comments that improved this paper.} This work was supported the NSF under award CAREER-1454333 and by NASA under \textit{Chandra} awards G07-18003A/B, GO7-18003A/B, GO8-9131X, and the ACIS Instrument Team contract SV4-74018; these were issued by the Chandra X-ray Center, which is operated by the Smithsonian Astrophysical Observatory for and on behalf of NASA under contract NAS8-03060. E.H.N. acknowledges prior support from the Cal-Bridge program through NSF award DUE-1356133. The scientific results are based in part on observations made by the \textit{Chandra X-ray Observatory} and published previously in cited articles. This work is based in part on archival data obtained with the \textit{Spitzer Space Telescope}, which is operated by the Jet Propulsion Laboratory, California Institute of Technology under a contract with NASA. This publication makes use of data products from the Two Micron All-Sky Survey, which is a joint project of the University of Massachusetts and the Infrared Processing and Analysis Center/California Institute of Technology, funded by NASA and the NSF.

\textit{Facility:} CXO (ACIS)

\textit{Facility:} Spitzer (IRAC)

\textit{Facility:} CTIO:2MASS

\bibliographystyle{aasjournal}
\bibliography{citations.bib}

\end{document}